# Welfare and Profit Maximization with Production Costs

Avrim Blum*     Anupam Gupta*     Yishay Mansour†     Ankit Sharma*

November 4, 2018

**Abstract**

Combinatorial Auctions are a central problem in Algorithmic Mechanism Design: pricing and allocating goods to buyers with complex preferences in order to maximize some desired objective (e.g., social welfare, revenue, or profit). The problem has been well-studied in the case of limited supply (one copy of each item), and in the case of digital goods (the seller can produce additional copies at no cost). Yet in the case of resources—oil, labor, computing cycles, etc.—neither of these abstractions is just right: additional supplies of these resources can be found, but at increasing difficulty (marginal cost) as resources are depleted.

In this work, we initiate the study of the algorithmic mechanism design problem of combinatorial pricing under increasing marginal cost. The goal is to sell these goods to buyers with unknown and arbitrary combinatorial valuation functions to maximize either the social welfare, or the seller's profit; specifically we focus on the setting of *posted item prices* with buyers arriving online. We give algorithms that achieve *constant factor* approximations for a class of natural cost functions—linear, low-degree polynomial, logarithmic—and that give logarithmic approximations for arbitrary increasing marginal cost functions (along with a necessary additive loss). We show that these bounds are essentially best possible for these settings.

*Computer Science Department, Carnegie Mellon University, Pittsburgh, PA 15213. This work was supported in part by the National Science Foundation under grants CCF-0830540 and CCF-1101215, as well as by the MSR-CMU Center for Computational Thinking.

†School of Computer Science, Tel Aviv University. This research was supported in part by the Google Inter-university center for Electronic Markets and Auctions, by a grant from the Israel Science Foundation, by a grant from United States-Israel Binational Science Foundation (BSF), and by a grant from the Israeli Ministry of Science (MoS).

# 1 Introduction

Combinatorial Auctions are a central problem in Algorithmic Mechanism Design: pricing and allocating goods to buyers with complex preferences in order to maximize some desired objective (e.g., social welfare, revenue, or profit). This problem is typically studied in one of two extreme cases – the case of *limited supply* (one copy of each item) or the case of *unlimited supply* (the seller can produce additional copies at no cost). For the case of limited supply, there are strong negative results (see Blumrosen and Nisan (2007) and the references therein) unless one makes additional assumptions on the buyers' valuations (e.g., submodularity (Dobzinski et al.; Lehmann et al., 2006; Dobzinski, 2007; Chakraborty et al., 2009)). In contrast, for unlimited supply, which is characteristic of digital goods, maximizing social welfare is trivial by giving all the items away for free, and for revenue maximization, good bounds can be achieved for general buyers (Briest et al., 2008; Balcan et al., 2008). However, in the case of resources—whether oil or computing cycles or food or attention span—the unlimited-supply case is too optimistic and the limited-supply case too pessimistic. More often than not, additional sources can be found, but at higher cost. Indeed, the classical market equilibrium in economic theory assumes that as prices rise, supply increases while demand decreases (giving a unique price which clears the market). Such a supply curve corresponds to a cost of obtaining goods that increases with the number of items desired.

In this work, we initiate the study of this setting where additional resources can be found, but at *increasing marginal cost*, for the algorithmic mechanism design problem of combinatorial pricing.[1] That is, a seller has $n$ goods, and for each good $i$ there is a non-decreasing marginal cost function $c_i()$, capturing the fact that additional units of this good can be obtained, but at an increasing difficulty to the seller per unit. We specifically focus on the most challenging setting of posted item prices[2] in the face of an unknown series of buyers, with unknown and arbitrary combinatorial valuation functions, who arrive online. That is, the seller (e.g., a supermarket) must assign prices to each of $n$ goods, then a buyer arrives with some arbitrary combinatorial valuation function and purchases the bundle maximizing her own quasilinear utility (valuation minus price). After the buyer has made her purchase, the seller may adjust prices, then the next buyer arrives, and so on. In this setting, the seller *cannot* ask a buyer to submit her utility function, cannot run VCG, and cannot charge an admission fee to enter the store. We consider two natural goals – maximizing social welfare (the sum of buyers' valuations on bundles purchased, minus the costs incurred by the seller for obtaining these items)[3] and maximizing profit (the total amount paid by the buyers, minus the costs incurred by the seller). Our main result is that using appropriate algorithms we can in fact perform nearly as well as in the much easier setting of digital goods for a wide range of cost curves.

A second scenario where our results are applicable is in the context of network routing with congestion. Specifically, we would like to maximize the sum of valuations of routed connections (each user has some pair of terminals and a valuation on receiving a connection) minus the congestion cost of routing them. The congestion cost can reflect either the energy required to support the traffic on the network or the cost of additional infrastructure needed to maintain the quality of service under increased load. Andrews et al. (2010) indicate that energy curves for processors exhibit dis-economies of scale i.e. energy expenditure is super-linear as a function of processor speed. Such a scenario is captured by our model of increasing marginal cost, which for network routing would mean increasing marginal congestion costs.

---
[1]One could also study decreasing marginal costs, though we point out that modeling decreasing marginal costs in a non-Bayesian adversarial setting poses difficulties. For example, there are situations where any algorithm can make positive profit only by initially going into deficit, at which point the adversary could send in no more buyers.

[2]By virtue of posted pricing, our mechanisms are inherently incentive compatible.

[3]Social welfare is a natural objective if we view the seller as a resource allocator within a company, and buyers as various units in the company needing resources.

We will sometimes refer to marginal cost to the seller of the $k^{th}$ copy of an item as the *production cost* of the $k^{th}$ copy of that item.

## 1.1 Our Results and Techniques

We focus on two goals: maximizing social welfare, and maximizing profit. Social welfare is the total valuation of the buyers for their bundles purchased, minus the cost to the seller of all items sold. That is, it is the total utility of all players including the seller. Note that because the production costs are not flat, even to maximize social welfare, one cannot simply sell items at their production costs; one must sell at a price higher than the production cost[4]. This is in order to ensure that items reach the buyers who want them (approximately) most. The second goal is to maximize profit, i.e., the sum of prices charged for items sold minus their costs to the seller.

For a wide range of reasonable cost functions (linear, low-degree polynomial, logarithmic), we present a pricing scheme that achieves a social welfare within a *constant* factor of the optimal social welfare allocation minus a necessary additive loss. This holds for buyers with arbitrary combinatorial valuation functions. Furthermore, the algorithm is quite 'natural' and reasonable: we price the $k^{th}$ copy of any good at the production cost of the $2k^{th}$ copy[5]. This pricing scheme, that we call *twice-the-index*, appears in Section 3.

The Twice-the-index pricing scheme however fails to give good guarantees for all increasing cost functions. For instance, for the 0-$\infty$ case, where the first few copies are available at zero cost and thereafter the copies have an extremely high production cost, buyer instances can be easily be created where twice-the-index fails to give any any 'reasonable' guarantee (Appendix A.2). Bartal et al. (2003) propose a pricing scheme for the 0-$\infty$ setting which achieves a logarithmic approximation to the social welfare in case the number of copies available at zero cost are logarithmically many. We build on their idea and apply it to an arbitrary increasing cost curve by breaking up the curve into contiguous chunks, each containing logarithmically many copies, and apply their pricing scheme separately for each chunk. This pricing scheme, presented in Section 4, gives roughly a logarithmic approximation to the optimal social welfare minus the production cost of logarithmically many initial copies.

While the Twice-the-index pricing scheme gives constant approximation guarantees for 'nice' curves, the pricing scheme in Section 4 gives a logarithmic approximation for arbitrary increasing curves. We would ideally want a single algorithm that can give us constant approximation guarantees for 'nice' curves and logarithmic guarantees for arbitrary increasing curves. We achieve this for the case of *convex* increasing curves. In Section 5, we present a *smoothing* pricing scheme that attains a constant approximation to the optimal social welfare for polynomial curves and a logarithmic approximation for arbitrary convex curves (plus some additive loss in both cases).

Interestingly, in order to prove the approximation guarantee for all of the presented social welfare maximizing schemes, we use a crucial result which we refer to as the *Structural Lemma*. This result is stated and proved in Section 2.1. The result reduces the problem of proving the social welfare guarantee of a pricing scheme for buyers with arbitrary valuations to a case of proving that for *every item*, the profit generated through sales of copies of the item is comparable to the area between the production curve of the item and a line parallel to $x$-axis and at a height equal to the prices of the lowest priced unsold copy of the item. Hence the Structural Lemma simplifies the analysis considerably since it allows the problem to be seen *per item* even though the original problem is combinatorial.

---

[4]See Appendix A.1.1 for an example.
[5]For illustrative examples showing why some closely related algorithms *fail*, see Appendix A.

Finally, in Section 6, we change our objective to maximizing the *profit*, i.e., the sum of prices of goods sold minus the production cost of the goods. Here we give a randomized pricing scheme that takes as input any social welfare maximizing scheme (with approximation factor, say $\rho$) and a single-buyer profit maximizing pricing (with approximation factor say $\mu$), and combines them to get a profit maximization pricing scheme that achieve a $(\rho + \mu)$ approximation to optimal profit for any sequence of buyers. In particular, we use the single-buyer profit maximization algorithm of Balcan et al. (2008) and combine it with the social-welfare pricing schemes mentioned above to get a logarithmic approximation to the optimal profit for arbitrary increasing curves. Our approach for combining a social welfare maximization pricing scheme with a single-buyer profit maximization algorithm is directly inspired by and builds upon a similar result presented in Awerbuch et al. (2003). In fact, it extends their results to a more general setting with production costs and arbitrary valuations.

## 1.2 Related work

There is a huge body of literature on combinatorial auctions and pricing algorithms: we refer the reader to (Blumrosen and Nisan, 2007; Hartline and Karlin, 2007) and the references therein—in particular, note (Bartal et al., 2003; Lehmann et al., 2006; Dobzinski et al.; Dobzinski, 2007; Briest et al., 2005; Lavi and Swamy, 2005). The setting of combinatorial auctions has been considered both in Bayesian (stochastic) settings, where the buyers' valuations are assumed to come from a known prior distribution, and non-Bayesian (adversarial) settings. Our work focuses on the non-Bayesian or adversarial setting.

The algorithms of Briest et al. (2005) give truthful mechanisms that achieve constant approximations to social welfare for $\Omega(\log n)$ copies of each item (see also (Archer et al., 2004)) in the *offline* setting. For the *online* setting, Bartal et al. (2003) give posted-price welfare-maximizing algorithms for combinatorial auctions in the limited supply setting—the approximation guarantees they give are logarithmic (when there are $\Omega(\log n)$ copies of each item) or worse (when there are fewer copies); their results are (nearly) tight for the online limited-supply setting. The smoothing algorithm presented in Section 5 generalizes the results of Bartal et al. (2003) to arbitrary increasing cost curves and not just 0-$\infty$ costs (i.e. the limited supply case).

The work of Awerbuch et al. (2003) shows how to convert deterministic (or some special kind of randomized) online mechanisms for allocation problems into (randomized) posted-pricing schemes that achieve $(\rho + \log V_{max})$-fraction of the optimal profit possible, where the online algorithm is $\rho$-competitive for the allocation problem and $V_{max}$ is the maximum valuation of any agent over the set of items. We extend their analysis to convert our social welfare maximizing algorithms to profit maximizing algorithms. The details of this conversion are given in Section 6.

## 2 Model, Notation, and Definitions

We consider the following setting. A seller is selling a set $\mathcal{I} = \{1, \ldots, n\}$ of $n$ items to a sequence $\mathcal{B}$ of $m$ buyers who arrive one at a time. The seller can obtain (or produce) additional copies of each item but at increasing (or at least non-decreasing) production cost; specifically, let $c_i(k)$ denote the production cost to the seller for the $k$th copy of item $i$. For each item $i$, let $C_i(k)$ be the cumulative cost for the first $k$ copies—i.e., $C_i(k) = \sum_{k' \leq k} c_i(k')$. Let $c_i^{\text{inv}}(p)$ be the number of copies of item $i$ available before the production cost exceeds $p$; in case $c_i(\cdot)$ is invertible, it follows that $c_i^{\text{inv}}(p) = c_i^{-1}(p)$.

Before each buyer arrives, the seller may mark up the costs to determine a sales price $\pi_i$ for each item $i$. Every buyer $b$ has some (unknown to the seller) valuation function $v_b : 2^\mathcal{I} \to \mathbb{R}$ over possible bundles of items (we only require that $v_b(\phi) = 0$ i.e. value on the empty bundle is zero), and purchases the utility-maximizing bundle for herself at the current prices. That is, buyer $b$ purchases the set $S$ maximizing $v_b(S) - \sum_{i \in S} \pi_i$. After a buyer finishes purchasing her desired set, the seller may then readjust prices, and then the next buyer arrives, and so on.

For any particular sequence of buyers, let opt be the allocation that maximizes the social welfare. Clearly, the social welfare achieved under opt, denoted by $W(\text{opt})$, is an upper bound on both the maximum social welfare and maximum profit achievable by any online algorithm.

For any algorithm alg, $W(\text{alg})$ shall denote the social welfare attained through the algorithm. The algorithm shall determine a pricing scheme for the seller and $\pi_i(k)$ shall denote the sales price charged for the $k^{th}$ copy of item $i \in \mathcal{I}$. While this could in principle depend on other items sold, for all our algorithms it will depend only on $k$ and the cost-curve for the item. $x_i$ shall denote the total number of copies of item $i$ sold by the algorithm, and $P_i^f$ shall denote the price of the first *unsold* copy of item $i$—i.e., $P_i^f = \pi_i(x_i + 1)$.

We shall denote the *total production cost* suffered by the algorithm by $C(\text{alg})$ and and the *revenue* made by $R(\text{alg})$. $\text{profit}_i$ shall denote the *profit* made by the algorithm from the sales of item $i$. The total profit made by the algorithm is $\sum_{i \in \mathcal{I}} \text{profit}_i = R(\text{alg}) - C(\text{alg})$.

Since $x_i$ are the total number of copies sold by the algorithm alg for item $i$, therefore, $C(\text{alg}) = \sum_{i \in \mathcal{I}} \sum_{k=1}^{x_i} c_i(k)$, $R(\text{alg}) = \sum_{i \in \mathcal{I}} \sum_{k=1}^{x_i} \pi_i(k)$ and $\text{profit}_i = \sum_{k=1}^{x_i} \pi_i(k) - \sum_{k=1}^{x_i} c_i(k)$.

The total valuation of buyers on their allocated bundles under alg is denoted by $V(\text{alg}) = \sum_{b \in \mathcal{B}} v_b(\text{alg}(b))$ where $\text{alg}(b)$ denotes the set of items bought by buyer $b$ from the algorithm alg. The *social welfare* made by the algorithm $W(\text{alg})$ is $V(\text{alg}) - C(\text{alg})$.

For opt, the welfare-maximizing allocation, $\lambda_i$ denotes the number of copies of item $i$ allocated in opt. $C(\text{opt})$, $V(\text{opt})$ and $W(\text{opt})$ are defined analogously.

### 2.1 Structural Lemma

A basic challenge for maximizing social welfare in the presence of increasing production costs is that if one charges too little, then items may be purchased by an initial sequence of buyers whose valuations are too low to generate much social welfare, until the production cost has jumped to a point where only very costly items remain that are out of reach of the subsequent high valuation buyers. On the other hand, if one charges too much, then one loses the opportunity to make certain sales. This problem is compounded by the fact that buyers may have very different combinatorial

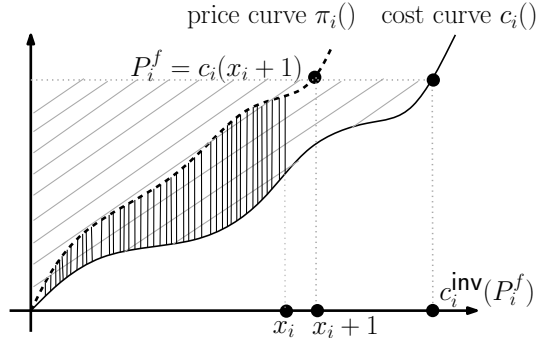

Figure 1: Structural Lemma: if the lightly shaded area is bounded by a small multiple of the doubly shaded area, then we get good social welfare. $x_i$ is the last sold copy of the item and $x_i + 1$ is the first unsold copy. The lower continuous curve is the cost curve while the upper dashed curve is the price curve.

preferences—one does not want to "run out" of cheap copies of one item for buyers who may have high valuation on large sets containing that item. In the following sections, we describe two pricing algorithms for addressing these issues and achieving good social welfare guarantees. In order to analyze the pricing algorithms, we first prove a key structural lemma regarding pricing under increasing production costs; this lemma will be used for all our subsequent analyses.

**Lemma 2.1.** *For a pricing algorithm alg with non-decreasing price functions $\pi_i$ suppose there exists some $\alpha \geq 1$ and $\beta \geq 0$ such that for every allowed set of values of the final prices $P_i^f$,*

$$\sum_{i \in \mathcal{I}} \sum_{k=1}^{c_i^{\mathsf{inv}}(P_i^f)} (P_i^f - c_i(k)) \leq \alpha \sum_{i \in \mathcal{I}} \mathsf{profit}_i + \beta ,\qquad(1)$$

*then on every instance of buyers*

$$W(\mathsf{alg}) \geq \tfrac{1}{\alpha}(W(\mathsf{opt}) - \beta) .$$

The term $\sum_{k=1}^{c_i^{\mathsf{inv}}(P_i^f)} (P_i^f - c_i(k))$ denotes the maximum possible social welfare which can be achieved by buyers who have valuation $P_i^f$ for item $i$ and zero for everything else. To see this note that (i) $P_i^f - c_i(k)$ is the contribution to social welfare if the $k^{th}$ of item $i$ is allocated to such a buyer and, (ii) the contribution $P_i^f - c_i(k)$ remains non-negative as long as $P_i^f \geq c_i(k)$ which is true only for $k \leq c_i^{\mathsf{inv}}(P_i^f)$. The theorem says that if for every possible set of final prices, we can bound such a social welfare summed over items by the profit generated by the algorithm, then for every sequence of buyers the algorithm gets a good social welfare compared to the optimum.

Graphically, as shown in Figure 1, $\sum_{k=1}^{c_i^{\mathsf{inv}}(P_i^f)} (P_i^f - c_i(k))$ is the area between the production curve $c_i()$ and the dotted line parallel to x-axis, marked by $P_i^f = c_i(x_i + 1)$, (the lightly shaded area) while $\mathsf{profit}_i$ is the region between the price curve and production curve (the doubly shaded area).

**Proof of Lemma 2.1 :** When buyer $b \in \mathcal{B}$ arrives, let $x_i^{(b)}$ be the number of copies of item $i$ sold before $b$ comes in. Hence, the price $b$ sees for item $i$ would be $\pi_i(x_i^{(b)} + 1)$; for brevity we denote this $q_b(i)$, and for a set $S \subseteq \mathcal{I}$, $q_b(S) := \sum_{i \in S} q_b(i)$. The utility of a set $S$ for buyer $b$ therefore is $v_b(S) - q_b(S)$. Since each buyer buys the set that maximizes her utility, hence in particular it implies that the set alg($b$) which buyer $b$ bought from alg must be giving her at least as much utility as the set opt allocated to her i.e.

$$v_b(S_b) - q_b(S_b) \geq v_b(S_b^*) - q_b(S_b^*).$$

Summing over all buyers, we get

$$\sum_{b \in \mathcal{B}}(v_b(S_b) - q_b(S_b)) \geq \sum_{b \in \mathcal{B}}(v_b(S_b^*) - q_b(S_b^*)).$$

Adding and subtracting $C(\mathsf{alg})$ and $C(\mathsf{opt})$ on the left hand and right hand sides respectively, we get

$$\Big(\sum_{b \in \mathcal{B}} v_b(S_b) - C(\mathsf{alg})\Big) - \Big(\sum_{b \in \mathcal{B}} q_b(S_b) - C(\mathsf{alg})\Big) \geq \Big(\sum_{b \in \mathcal{B}} v_b(S_b^*) - C(\mathsf{opt})\Big) - \Big(\sum_{b \in \mathcal{B}} q_b(S_b^*) - C(\mathsf{opt})\Big).$$

Identifying the term $\sum_{b \in \mathcal{B}} v_b(S_b) - C(\mathsf{alg})$ with $W(\mathsf{alg})$, the term $\sum_{b \in \mathcal{B}} q_b(S_b) - C(\mathsf{alg})$ with $\sum_{i \in \mathcal{I}} \mathsf{profit}_i$ and the term $\sum_{b \in \mathcal{B}} v_b(S_b^*) - C(\mathsf{opt})$ with $W(\mathsf{opt})$ we get

$$W(\mathsf{alg}) - \sum_{i \in \mathcal{I}} \mathsf{profit}_i \geq W(\mathsf{opt}) - \Big(\sum_{b \in \mathcal{B}} q_b(S_b^*) - C(\mathsf{opt})\Big). \qquad (2)$$

Since prices are non-decreasing, hence the price faced by any buyer cannot be more than the final price of the various items. Therefore for each buyer $b$, $q_b(S_b^*) = \sum_{i \in S_b^*} \pi_i(x_i^{(b)} + 1) \leq \sum_{i \in S_b^*} \pi_i(x_i + 1) = \sum_{i \in S_b^*} P_i^f$. Hence, the term $\sum_{b \in \mathcal{B}} q_b(S_b^*)$ is at most $\sum_{b \in \mathcal{B}} \sum_{i \in \mathsf{opt}(b)} P_i^f = \sum_{i \in \mathcal{I}} (P_i^f \cdot \lambda_i)$ where recall that $\lambda_i$ denotes the number of copies of item $i$ allocated under opt. Moreover, since $C(\mathsf{opt}) = \sum_{i \in \mathcal{I}} \sum_{k=1}^{\lambda_i} c_i(k)$, we have

$$\sum_{b \in \mathcal{B}} q_b(S_b^*) - C(\mathsf{opt}) \leq \sum_{i \in \mathcal{I}} (P_i^f \cdot \lambda_i) - \sum_{i \in \mathcal{I}} \sum_{k=1}^{\lambda_i} c_i(k)$$
$$= \sum_{i \in \mathcal{I}} \sum_{k=1}^{\lambda_i} (P_i^f - c_i(k)). \qquad (3)$$

The quantity $(P_i^f - c_i(k))$ is non-negative until $c_i(k) \leq P_i^f$, that is it is non negative for $k \leq c_i^{\mathsf{inv}}(P_i^f)$. Hence, we have $\sum_{b \in \mathcal{B}} q_b(S_b^*) - C(\mathsf{opt}) \leq \sum_{i \in \mathcal{I}} \sum_{k=1}^{\lambda_i} (P_i^f - c_i(k)) \leq \sum_{i \in \mathcal{I}} \sum_{k=1}^{c_i^{\mathsf{inv}}(P_i^f)} (P_i^f - c_i(k))$. Therefore using Equation (2) we get

$$W(\mathsf{alg}) - \sum_{i \in \mathcal{I}} \mathsf{profit}_i \geq W(\mathsf{opt}) - \Big(\sum_{b \in \mathcal{B}} q_b(S_b^*) - C(\mathsf{opt})\Big) \geq W(\mathsf{opt}) - \sum_{i \in \mathcal{I}} \sum_{k=1}^{c_i^{\mathsf{inv}}(P_i^f)} (P_i^f - c_i(k)).$$

If $\sum_{i \in \mathcal{I}} \sum_{k=1}^{c_i^{\mathsf{inv}}(P_i^f)} (P_i^f - c_i(k)) \leq \alpha \sum_{i \in \mathcal{I}} \mathsf{profit}_i + \beta$, then using above equation we get

$$W(\mathsf{alg}) - \sum_{i \in \mathcal{I}} \mathsf{profit}_i \geq W(\mathsf{opt}) - (\alpha \sum_{i \in \mathcal{I}} \mathsf{profit}_i + \beta) \Rightarrow W(\mathsf{alg}) + (\alpha - 1) \sum_{i \in \mathcal{I}} \mathsf{profit}_i \geq W(\mathsf{opt}) - \beta.$$

Finally using the social welfare generated by the algorithm is at least the profit made, i.e. $W(\mathsf{alg}) \geq \sum_{i \in \mathcal{I}} \mathsf{profit}_i$, we get the desired result $W(\mathsf{alg}) \geq (W(\mathsf{opt}) - \beta)/\alpha$ □

In Section B we present a variant of the structural lemma that will be useful for the analysis of the pricing algorithms presented in Section 4 and Section 5. In the following section, we give pricing strategies that satisfy Lemma 2.1 (or its variant) for suitable $\alpha, \beta$.

# 3 Algorithm: Pricing at twice the index

The first two ideas for pricing items with production costs are perhaps to (a) sell at cost, or (b) sell at some constant times the cost; however, these schemes fail even for simple cost functions like linear and logarithmic production costs, respectively. (See Appendix A for some examples.) In this section, we consider the next natural pricing scheme: *The price $\pi_i(k)$ of the $k^{th}$ copy of an item is the production cost of the $(2k)^{th}$ copy.* I.e.,

$$\pi_i(k) := c_i(2k).$$

There is nothing special about pricing at *twice* the index, other factors would work as well, just giving slightly different bounds. We shall analyze this algorithm for function classes including polynomial $c_i(x) = x^d$ and logarithmic $c_i(x) = \ln(1+x)$. Since these functions are strictly increasing and hence invertible, hence we shall have $c_i^{\mathsf{inv}}(c_i(x)) = x$ for all $x \geq 0$. To analyze this algorithm, we shall use the result of Lemma 2.1.

Define $A_i(x_i) := \sum_{k=1}^{c_i^{\mathsf{inv}}(P_i^f)} (P_i^f - c_i(k))$. To apply Lemma 2.1, we will show that $\forall x_i \geq 0, A_i(x_i) \leq \alpha \cdot \mathsf{profit}_i(x_i) + \beta_i$ and thereby get $\sum_{i \in \mathcal{I}} A_i(x_i) \leq \alpha \sum_{i \in \mathcal{I}} \mathsf{profit}_i(x_i) + \beta$ where $\beta = \sum_{i \in \mathcal{I}} \beta_i$.

Since the price $\pi_i(k)$ of the $k^{th}$ copy is $c_i(2k)$, hence the profit made from the sales of such of a copy is $c_i(2k) - c_i(k)$. Further, since $x_i$ copies of item $i$ have been sold, therefore, $P_i^f = c_i(2(x_i+1))$ and hence $c_i^{\mathsf{inv}}(P_i^f) = 2x_i + 1$. Therefore, when pricing at twice the index, we have $A_i(x_i) = \sum_{k=1}^{2(x_i+1)} (c_i(2(x_i+1)) - c_i(k))$ and $\mathsf{profit}_i(x_i) = \sum_{k=1}^{x_i} (c_i(2k) - c_i(k))$.

## 3.1 Performance on some cost functions

We now show that for some "well-behaved" classes of functions, we get $A_i(x) \leq \alpha \cdot \mathsf{profit}_i(x) + \beta_i$; the $\beta_i$ term will usually depend on the production cost of the first few copies of the items—hence we will guarantee that we get a multiplicative $\alpha$-fraction of the welfare if we ignore the production cost of the first few copies.

- Linear production costs: $c_i(x) = a_i x + b_i$ for some constant $a_i, b_i \geq 0$, then we have $A_i(x) = a_i(x+1)(2x+1)$, and $\mathsf{profit}_i(x) = \frac{1}{2} a_i x(x+1)$, and hence $A_i(x) \leq 6\,\mathsf{profit}_i + a_i$. Lemma 2.1 implies that

$$\begin{aligned} W(\mathsf{alg}) &\geq \tfrac{1}{6}\left(W(\mathsf{opt}) - \sum_{i \in \mathcal{I}} a_i\right) \\ &= \tfrac{1}{6}\left(W(\mathsf{opt}) - \sum_{i \in \mathcal{I}} (c_i(2) - c_i(1))\right). \end{aligned}$$

This result, with suitably modified guarantees, can easily be extended to the case where the actual production cost lies between *two linear curves* whose slopes are within a constant factor of each other.

- Polynomial production costs: $c_i(x) = a_i x^d$ for $d > 1$. Then $A_i(x) \leq a_i \frac{d}{d+1} (2(x+1))^{d+1}$, whereas $\mathsf{profit}_i(x) \geq a_i \frac{1}{d+1}(2^d - 1) x^{d+1}$, so some algebra implies that $A_i(x) \leq 12\,d\,\mathsf{profit}_i(x) + 2^{d+1}(d+2)^{d+1} a_i$. Hence

$$W(\mathsf{alg}) \geq \tfrac{1}{12d}\left(W(\mathsf{opt}) - 2(d+2)^{d+1} \sum_{i \in \mathcal{I}} c_i(2)\right).$$

Such a bound also holds for $c_i(x)$ being a polynomial of degree at most $d$ with positive coefficients. The additive loss of $2^{O(d \log(d))}$ should be compared to the lower bound of $\Omega(2^d/d)$ in Corollary A.3

- Logarithmic production costs: $c_i(x) = \ln(1+x)$. By algebra, $A_i(x) \leq (2x+3)$, and $\mathsf{profit}_i(x) \geq \ln(\frac{3}{2}) x$, so $A_i(x) \leq \frac{2}{\ln(3/2)} \mathsf{profit}_i(x) + 3$, and Lemma 2.1 implies

$$W(\mathsf{alg}) \geq \tfrac{\ln(3/2)}{2} \left(W(\mathsf{opt}) - 3|\mathcal{I}|\right) .$$

## 3.2 Trade-off between the multiplicative guarantee and additive loss

In the guarantees given above, gains in the multiplicative factor can be made while trading-off commensurate losses in the additive loss terms. Specifically, consider the polynomial production cost $c_i(x) = x^d$. For a given $x_i$, we have that $A_i(x_i) \leq \frac{d}{d+1} (2(x_i+1))^{d+1}$ and $\mathsf{profit}_i(x_i) \geq \frac{(2^d-1) x_i^{d+1}}{d+1}$. Hence,

$$A_i(x_i) \leq \frac{d}{d+1} (2(x_i+1))^{d+1} = \frac{d}{d+1} (1+1/x_i)^{d+1} 2^{d+1} x_i^{d+1} \leq 4\, d\, (1+1/x_i)^{d+1} \mathsf{profit}_i(x_i) \quad (4)$$

where we have used $\forall d \geq 1,\ 2^d - 1 \geq 2^{d-1}$. Therefore, using that

- for all $x_i \leq q$, $A_i(x_i) \leq \frac{d}{d+1} (2(x_i+1))^{d+1} \leq \frac{d}{d+1} (2(q+1))^{d+1}$, and
- for all $x_i > q$, $A_i(x_i) \leq 4\, d\, (1+1/x_i)^{d+1} \mathsf{profit}_i \leq 4\, d\, (1+1/q)^{d+1} B_i(x_i)$,

for any $q \geq 1$, we can write

$$\forall x_i \geq 0,\ A_i(x_i) \leq 4\, d\, (1+1/q)^{d+1} \mathsf{profit}_i(x_i) + (d/(d+1)) (2(q+1))^{d+1} .$$

Denoting $\alpha(q) = 4\, d\, (1+1/q)^{d+1}$ and $\beta(q) = (d/(d+1)) (2(q+1))^{d+1}$ we can write for any $q \geq 1$ using Lemma 2.1, $W(\mathsf{alg}) \geq (W(\mathsf{opt}) - \sum_{i \in \mathcal{I}} \beta(q))/\alpha(q)$. A large $q$ means a higher additive loss but with the benefit of a lower multiplicative factor. Hence, depending on the specific situation, we can look for a sweet spot by varying the parameter $q$. In the previous section, we had chosen $q = d+1$ to give the result for polynomial case.

As we show in Appendix A.3, a social-welfare maximizing algorithm which has no estimate of $W(\mathsf{opt})$ has to lose an additive factor. At a high level, $q$ represents the number of initial copies which we are ready to lose.

While the "twice-the-index" algorithm works for the above cost functions, its behavior worsens if the function grows very fast; Appendix A.2 shows a bad example for the algorithm. Hence, in the next section, we give a logarithmic-approximation algorithm for the case of arbitrary increasing production cost curves.

## 4 Arbitrary Increasing Cost Functions

In this section, we present an algorithm that applies to arbitrary increasing cost functions, giving a logarithmic approximation minus an additive term that depends on the cost function (Theorem 4.2). The guarantee is achieved through a simple discretization of the cost function that allows us to reduce to the case of step functions and apply the algorithm of Bartal et al. (2003). In fact, we get a multiplicative logarithmic approximation to $W(\mathsf{opt})$ as long as the production cost of the first few logarithmic copies of all the items is small compared to $W(\mathsf{opt})$. For the $0 - \infty$ production cost setting (i.e. the first few copies at zero cost and subsequent at an extremely high cost), if we have $\Omega(\log nm)$ copies of each item available at zero cost, the additive loss is zero and the algorithm presented here gets a logarithmic fraction of the optimal social welfare just as in Bartal et al. (2003).

## 4.1 Algorithm

Before describing the algorithm, let us introduce some notation. Define $U_{max}$ as the maximum welfare any single buyer can achieve. Mathematically,

$$U_{max}(\mathcal{I}, \mathcal{B}) = \max_{b \in \mathcal{B}} \max_{T \subseteq \mathcal{I}} \left( v_b(T) - \sum_{i \in T} c_i(1) \right), \tag{5}$$

Note that the optimal social welfare, $W(\mathsf{opt})$, lies between $U_{max}$ and $m \cdot U_{max}$. The algorithm requires a parameter $Z$ which satisfies $Z \in (U_{max}, U_{max}/\epsilon]$ [6]. For item $i$, define $\ell_i = \min\{c_i^{\mathsf{inv}}(Z), m\}$ and $c_i^{\mathsf{invt}}(p) = \min\{c_i^{\mathsf{inv}}(p), c_i^{\mathsf{inv}}(Z), m\}$. We can think of $\ell_i$ as the 'effective' number of copies of item $i$ that are available and of $c_i^{\mathsf{invt}}(p)$ as the function which gives the 'effective' number of copies of item $i$ whose production cost is at most $p$; $c_i^{\mathsf{invt}}(p)$ is the maximum number of copies of item $i$ that $\mathsf{opt}$ can allocate before the production cost exceeds $p$ (Corollary B.1). Note that using $c_i^{\mathsf{invt}}$ (as opposed to using $c_i^{\mathsf{inv}}$) is a technicality; one can imagine $c_i^{\mathsf{invt}} \approx c_i^{\mathsf{inv}}$ for a first read.

We now describe the pricing algorithm. In order to price copies for an item $i$, the algorithm divides $\ell_i$ copies into contiguous steps and each step has $\tau_i$ number of copies where $\tau_i = \lceil \log(4 n \ell_i/\epsilon) \rceil$; hence the first step contains copies 1 through $\tau_i$, the second from $\tau_i + 1$ through $2\tau_i$ and so on. Let $s_{rq}$ denote the $q^{th}$ copy relative to the $r^{th}$ step; note that $q$ varies from 1 to $\tau$. The production cost of copy $s_{rq}$ is therefore $c_i((r-1) \cdot \tau_i + q)$; the first copy in step $r$ has cost $c_i((r-1)\tau_i + 1)$ and the last copy has cost $c_i(r \cdot \tau_i)$.

The algorithm sets the price of copy $s_{rq}$ as

$$\pi_i(s_{rq}) = \frac{\epsilon Z}{4 n \ell_i} \cdot 2^q + c_i(r \cdot \tau_i)$$

so that the first copy in step $r$ has price $\frac{\epsilon Z}{4 n \ell_i} + c_i(r \cdot \tau_i)$ while the last copy has price at least $Z + c_i(r \cdot \tau_i)$. Note that since any copy in the $r^{th}$ step has production cost at most $c_i(r \cdot \tau_i)$, therefore, the price of every copy in the $r^{th}$ step is greater than its production cost.

For every item, the algorithm sells copies of the item in increasing order of prices, so it might so happen that after the sale of a few copies from the first step, copies from the second step start selling, even before all copies of the first step are exhausted, since the copies in the second step are cheaper than the copies remaining in the first step.

## 4.2 Analysis

The crucial lemma of this section that will help prove the social welfare guarantee is

**Lemma 4.1.** *For every item $i \in \mathcal{I}$, $\sum_{k=1}^{c_i^{\mathsf{invt}}(P_i^f)} (P_i^f - c_i(k)) \leq 4 \cdot \tau_i \cdot \mathsf{profit}_i + \frac{\epsilon Z}{2 n} + (c_i(\tau_i) - c_i(1)) \cdot \tau_i$*

We now use Lemma 4.1 to prove the main result of this section.

---
[6] We can remove this assumption at a further loss of $O(\log W(\mathsf{opt}) (\log \log W(\mathsf{opt}))^2)$ in the approximation guarantee (Balcan et al., 2008).

**Theorem 4.2.** *Given a parameter $Z \in (U_{max}, U_{max}/\epsilon]$, the social welfare $W(\text{alg})$ achieved by the algorithm satisfies*

$$W(\text{alg}) \geq \frac{W(\text{opt})/2 - \sum_{i \in \mathcal{I}}(c_i(\tau_i) - c_i(1)) \cdot \tau_i}{4 \cdot \max_{i \in \mathcal{I}} \tau_i}$$

*where $\tau_i = \lceil \log(4\, n\, \ell_i/\epsilon) \rceil$ and $\ell_i = \min\{c_i^{\text{inv}}(Z), m\}$.*

Roughly, Theorem 4.2 states that the social welfare achieved by the algorithm is a logarithmic approximation to the optimal social welfare minus the sum of production cost of the first few copies of every item.

**Proof of Theorem 4.2:** Combining the result of Lemma 4.1 over all items $i \in \mathcal{I}$, we get

$$\sum_{i \in \mathcal{I}} \sum_{k=1}^{c_i^{\text{invt}}(P_i^f)} (P_i^f - c_i(k)) \leq 4 \cdot \max_{i \in \mathcal{I}} \tau_i \cdot \sum_{i \in \mathcal{I}} \text{profit}_i + \frac{\epsilon Z}{2} + \sum_{i \in \mathcal{I}}(c_i(\tau_i) - c_i(1)) \cdot \tau_i \ .$$

We now use the variant of structural lemma, Corollary B.1, stated in Appendix B, to get

$$W(\text{alg}) \geq \frac{W(\text{opt}) - \frac{\epsilon Z}{2} - \sum_{i \in \mathcal{I}}(c_i(\tau_i) - c_i(1)) \cdot \tau_i}{4 \cdot \max_{i \in \mathcal{I}} \tau_i}$$

and finally use $W(\text{opt}) - \frac{\epsilon Z}{2} \geq W(\text{opt})/2$ (which is implied by $Z \in (U_{max}, U_{max}/\epsilon]$) to get the desired result. □

We now need to prove Lemma 4.1. The analysis below considers any particular item $i \in \mathcal{I}$. Recall that $P_i^f$ denotes the price of the lowest price unsold copy of item $i$. Let $t$ be the step which contains the copy $c_i^{\text{invt}}(P_i^f)$. Define for $1 \leq r < t$, $s_r = \tau_i$, and $s_t = \min\{\tau_i, c_i^{\text{invt}}(P_i^f) - (t-1)\tau_i\}$ so that we have $\sum_{r=1}^{t} s_r = c_i^{\text{invt}}(P_i^f)$. Further, for item $i$, let $\text{profit}_i(r)$ denote the total profit made by the algorithm from the sales of copies of the item from its $r^{th}$ step. Finally for convenience of analysis define $c_i(0) = c_i(1)$.

The following lemma bounds the left hand side of the inequality claimed in Lemma 4.1 in terms of a related quantity.

**Lemma 4.3.** $\sum_{k=1}^{c_i^{\text{invt}}(P_i^f)} (P_i^f - c_i(k)) \leq \sum_{r=1}^{t}(P_i^f - c_i((r-1) \cdot \tau_i)) \cdot s_r$

*Proof.* Note that $\sum_{k=1}^{c_i^{\text{invt}}(P_i^f)}(P_i^f - c_i(k)) = \sum_{r=1}^{t} \sum_{x=1}^{s_r}(P_i^f - c_i((r-1) \cdot \tau_i + x))$ where we have broken up the summation across the different steps. Finally, $c_i((r-1) \cdot \tau_i + x) \geq c_i((r-1) \cdot \tau_i)$ since we are dealing with a non-decreasing production curve $c_i()$ and therefore for each $r$, we have $\sum_{x=1}^{s_r}(P_i^f - c_i((r-1) \cdot \tau_i + x)) \leq (P_i^f - c_i((r-1) \cdot \tau_i)) \cdot s_r$. This gives us the desired result. □

**Lemma 4.4.** *For each step $r$ such that $2 \leq r \leq t$, $(P_i^f - c_i(r \cdot \tau_i)) \leq 2 \cdot \text{profit}_i(r) + \frac{\epsilon Z}{4\, n\, \ell_i}$.*

*Proof.* This is because

- either $(P_i^f - c_i(r \cdot \tau_i)) > \frac{\epsilon Z}{4\, n\, \ell_i}$, in which case $\text{profit}_i(r) \geq (P_i^f - c_i(r \cdot \tau_i))/2$.

  This is because for every $p$ such that $\frac{\epsilon Z}{4\, n\, \ell_i} \leq p \leq Z$, the $r^{th}$ step has a copy whose price is in the range $[p/2 + c_i(r \cdot \tau_i), p + c_i(r \cdot \tau_i))$ and hence in particular, there is a copy in the $r^{th}$ step whose price $q$ is in the range $[(P_i^f - c_i(r \cdot \tau_i))/2 + c_i(r \cdot \tau_i), (P_i^f - c_i(r \cdot \tau_i)) + c_i(r \cdot \tau_i)) = [(P_i^f - c_i(r \cdot \tau_i))/2 + c_i(r \cdot \tau_i), P_i^f)$. Therefore the price $q$ of such a copy is strictly less than $P_i^f$ and since the $P_i^f$ is the price of the lowest priced unsold copy of item $i$, therefore the copy at price $q$ must have been sold. Any copy in $r^{th}$ step has production cost at most $c_i(r \cdot \tau_i)$, hence the sale of a copy at price $q \geq (P_i^f - c_i(r \cdot \tau_i))/2 + c_i(r \cdot \tau_i)$ must result in a profit of at least $(P_i^f - c_i(r \cdot \tau_i))/2$.

- or $(P_i^f - c_i(r \cdot \tau_i)) \leq \frac{\epsilon Z}{4n\ell_i}$.

Since $\mathsf{profit}_i(r)$ is a non-negative quantity, hence we see that in both cases the desired inequality is satisfied. □

**Proof of Lemma 4.1 :**

Note that

$$\sum_{r=1}^{t}(P_i^f - c_i((r-1) \cdot \tau_i)) \cdot s_r = (P_i^f - c_i(0)) \cdot s_1 + \sum_{r=2}^{t}(P_i^f - c_i((r-1) \cdot \tau_i)) \cdot s_r \qquad (6)$$

First, using Lemma 4.4 we bound the second summation on the right hand side of equation (6).

$$\sum_{r=2}^{t}(P_i^f - c_i((r-1) \cdot \tau_i)) \cdot s_r \leq 2 \cdot \sum_{r=2}^{t} \mathsf{profit}_i(r-1) \cdot s_r + \sum_{r=2}^{t} \frac{\epsilon Z}{4n\ell_i} \cdot s_r$$

$$\leq 2 \cdot (\max_r s_r) \cdot \sum_{r=2}^{t} \mathsf{profit}_i(r-1) + \frac{\epsilon Z}{4n\ell_i} \cdot \sum_{r=2}^{t} s_r$$

$$\leq 2 \cdot \tau_i \cdot \mathsf{profit}_i + \frac{\epsilon Z}{4n} \qquad (7)$$

where in the last inequality we have used $\sum_{r=2}^{t} s_r = c_i^{\mathsf{invt}}(P_i^f) \leq \ell_i$ and that $\tau_i \geq s_r$ for any $r$.

Now we bound the first term on the right hand side of equation (6).

$$(P_i^f - c_i(0)) \cdot s_1 = (P_i^f - c_i(\tau_i)) \cdot s_1 + (c_i(\tau_i) - c_i(0)) \cdot s_1$$

$$\leq 2 \cdot \tau_i \cdot \mathsf{profit}_i(1) + \frac{\epsilon Z}{4n} + (c_i(\tau_i) - c_i(0)) \cdot s_1$$

$$\leq 2 \cdot \tau_i \cdot \mathsf{profit}_i + \frac{\epsilon Z}{4n} + (c_i(\tau_i) - c_i(0)) \cdot s_1 \qquad (8)$$

where the first inequality follows from Lemma 4.4 and the second follows from noting that the total profit $\mathsf{profit}_i$ made through sales of copies of item $i$ is at least as much as the profit $\mathsf{profit}_i(1)$ made through the sale of copies from the first step of the item.

Using Lemma 4.3 and Equations (6), (7) and (8) derived above we get

$$\sum_{k=1}^{c_i^{\mathsf{invt}}(P_i^f)} (P_i^f - c_i(k)) \leq 4 \cdot \tau_i \cdot \mathsf{profit}_i + 2 \cdot \frac{\epsilon Z}{4n} + (c_i(\tau_i) - c_i(0)) \cdot \tau_i$$

Using Claim 4.3 and noting that by definition $c_i(0) = c_i(1)$, we get the desired result. □

# 5 Smoothing Algorithm

The pricing algorithm of Section 4 gives us a logarithmic multiplicative guarantee along with some additive loss for all increasing cost curves. Twice-the-index algorithm presented in Section 3 gives a constant approximation factor plus an additive loss for polynomial curves. This raises the question of whether there is a pricing algorithm which can achieve the best of both the worlds i.e. give logarithmic multiplicative guarantees for general curves but constant factor guarantee for nice curves such as polynomial and logarithmic. In this section we present a pricing algorithm that achieves that for the case of *convex* cost functions. It gives logarithmic guarantees for general convex curves (Corollary 5.6) but in addition, gives for polynomial cost curves, a constant factor approximation (Theorem 5.15).

## 5.1 Intuition

Ideally, we would like to set prices which are sufficiently far above the cost curve (so that we generate a large social welfare), yet not be too far above it (else the high prices may result in no sales, causing a large additive loss). Hence, we run into problems when the cost curve increases sharply—and the intuitive goal is to create a price curve which smooths out these sharp changes in the cost curve while staying "close" to it.

The smoothing algorithm takes the cost curve, and creates a price function which is a monotone step function: copies of the item are grouped into intervals, with all copies in an interval having the same price. We call these intervals "price intervals". The algorithm creates the price curve from *right to left.* If we think of $\ell_i$ as the effective number of copies of item $i$ and $Z$ as the highest price, then the $\ell_i^{th}$ copy is priced first at price $Z$ through creation of the price interval $[\lfloor \frac{2}{3}\ell_i \rfloor, \infty)$[7], with items in this interval priced at $Z$; subsequently, price intervals are created progressively moving leftwards until we have priced the first copy. At each point, we use the intuition from above: if the price is much higher than the cost, we set the price for the new interval such that the price-cost gap is slashed by a factor of 2, else we set the price to maintain a sufficient gap from the cost.

## 5.2 The smoothing algorithm

Before we give the algorithm (in Figure 2), let us give some definitions; we urge the impatient reader to jump to Section 5.3 to get a quick rough feel of the algorithm. We assume that the cost of the first copy of every item is 0 i.e. $\forall i, c_i(1) = 0$ [8]. Recall the notation $U_{max}$ defined in Equation 5 in Section 4; it represented the maximum welfare which can be made through a single buyer. In the present scenario since $c_i(1) = 0$, hence $U_{max}$ equals $\max_{b \in \mathcal{B}} \max_{T \subseteq \mathcal{I}} v_b(T)$

Define $\ell_i = \min\{c_i^{\mathsf{inv}}(Z), m\}$ and $B_i = \lceil 12 \log(4n\ell_i/\epsilon) \rceil$. Similar to Section 4, at a high level, think of $\ell_i$ as being the "effective number" of copies of item $i$ available, and $B_i$ as the "number of different price levels" we create in our price curve. $c_i^{\mathsf{invt}}(p)$, as in Section 4, is defined as the "truncated" value $\min\{c_i^{\mathsf{inv}}(p), c_i^{\mathsf{inv}}(Z), m\}$; please refer to Section 4 to get a sense of why $c_i^{\mathsf{invt}}$ is defined the way it is. Define $\mathsf{width}_i(p) := \lfloor \frac{c_i^{\mathsf{invt}}(p)}{B_i} \rfloor$; this function will determine the number of copies we group together in a price interval. We assume that

$$\ell_i \geq B_i \geq 12; \qquad (9)$$

---

[7] We abuse notation slightly by denoting the integer interval $\{r, r+1, \ldots, s-1\}$ as the half-open real interval $[r, s)$.
[8] In Lemma D.1 we show that this is without loss of generality.

see Claim C.3 for why this is without loss of generality.

---

1: **for all** $x \geq \lfloor \frac{2}{3} \ell_i \rfloor$, **set** $\pi_i(x) := Z$
2: **set** $x \leftarrow \lfloor \frac{2}{3} \ell_i \rfloor$
3: **while** $x > 1$ **do**
4:   **if** $\mathsf{width}_i(\pi_i(x)) \geq 1$ **then**
5:     **set** $x' \leftarrow \max\{x - \mathsf{width}_i(\pi_i(x)), 1\}$
6:

$$\mathbf{set}\ \Delta = \begin{cases} \frac{\pi_i(x) - c_i(x)}{2} & \text{if } \pi_i(x) \geq 3\, c_i(x) \\ \frac{c_i(x)}{2} & \text{otherwise} \end{cases}$$

7:     for all $y \in [x', x)$, **set** $\pi_i(y) := c_i(x) + \Delta$
8:     **set** $x \leftarrow x'$
9:   **else**
10:     for all $y \in [1, x)$, **set** $\pi_i(y) := \pi_i(x)$
11:     **set** $x \leftarrow 1$

---

Figure 2: Smoothing algorithm

Let $\pi_i : \mathbb{Z}_+ \to \mathbb{R}_+$ be the price function and let $\mathcal{J}_i$ denote the set of price intervals for item $i$, and with $z_i = |\mathcal{J}_i|$. We refer to the $q^{th}$ interval of item $i$ as $J_{iq}$, with $J_{i1}$ being the price interval that contains the first copy of item $i$, and $J_{iz_i} = [\lfloor \frac{2}{3} \ell_i \rfloor, \infty)$. Let $\pi_i(J_{iq})$ be the price of the copies in the interval $J_{iq}$. Depending on the production curve, two consecutive price intervals may have the same price. Also, we will formally state later that the prices we generate are non-decreasing, and always stay above the production cost for all copies less than $\ell_i$.

## 5.3 The main ideas

**Smoothing:** Step 6 ensures a smooth price curve: if the price is more than thrice the production cost, we slash the gap between the price and production cost by two else we allow the price to stay at a sufficient gap from the cost.

**Price Interval Size :** The idea of the analysis is to show that whenever the number of copies sold moves from a lower price interval to a higher one, the social welfare generated by selling copies at the lower price is enough to be competitive against opt, even if we sell no further copies at the higher price. Consequently, the size of a price interval $J_{iq}$ must depend on the price of items in the next interval $J_{iq+1}$. It turns out that to get a multiplicative approximation factor of $O(B_i)$, if the price of copies in $J_{iq+1}$ were $P$, it suffices to set the width of $J_{iq}$ to be $\lfloor \frac{c_i^{\mathsf{invt}}(P)}{B_i} \rfloor = \mathsf{width}_i(P)$. Here is a simple special case that illustrates why: suppose only item $i$ was being sold and we sold all copies from $J_{iq}$ but no copies from interval $J_{iq+1}$. We would like to apply Lemma 5.4. The final price $P_i^f$ in that case is $P = \pi_i(J_{iq+1})$. Staring at the left hand side of Equation (1), we see that it is at most $P \cdot c_i^{\mathsf{invt}}(P)$. Since we sold all the copies in price interval $J_{iq}$, we sold at least $|J_{iq}| = \lfloor \frac{c_i^{\mathsf{invt}}(P)}{B_i} \rfloor$ many copies, each at profit at least $P/6$ (something we will prove later). Hence on the right hand side of Equation (1), the term $\mathsf{profit}_i$ is at least $P \cdot \lfloor \frac{c_i^{\mathsf{invt}}(P)}{B_i} \rfloor / 6$. Putting $\alpha = O(B_i)$ we satisfy Equation (1) and thereby get an $O(B_i)$ approximation. Since the width of $J_{iq}$ depends on the price of $J_{i\,q+1}$, it is natural that our pricing algorithm creates price intervals from *right to left*.

**Termination:** The algorithm terminates in one of two ways: either while creating such appropriately sized price intervals, we hit the first copy (i.e., $x' \leftarrow 1$ in Step 5, and then the loop condition fails in Step 3) or the price $p$ of some price interval is low enough that $p < c_i(B_i)$, which implies $c_i^{\mathsf{invt}}(p) < B_i$ (the proof of implication appears later) and therefore $\mathsf{width}_i(p) = \lfloor \frac{c_i^{\mathsf{invt}}(p)}{B_i} \rfloor < 1$: this causes $x \leftarrow 1$ in Step 11. In the latter case, the price has become low enough that we can simply group all remaining copies into the lowest priced interval $J_{i1}$ at price $p$. The subsequent analysis will often have to separately consider these two cases: whether $x \leftarrow 1$ is achieved in Step 5 or in Step 11.

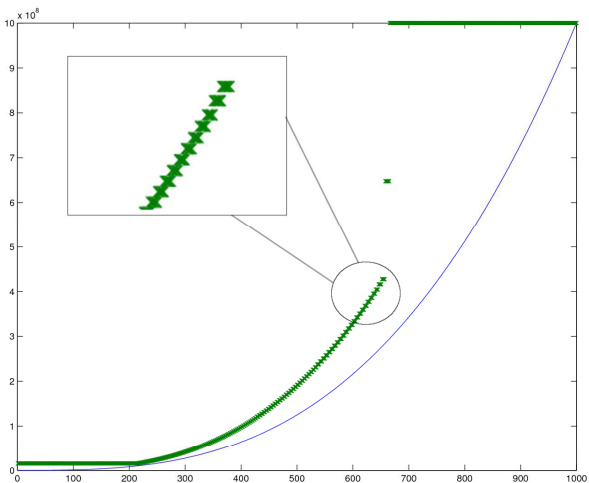

Figure 3: The figure shows the pricing curve drawn by the smoothing algorithm for the production curve $c_i(x) = x^3$. The lower line is the production curve. The upper thicker line is the pricing curve. We can observe that the price curve is flat towards the extreme right; this flat region contains the right-most price interval. Towards the extreme left the price curve *appears* to be a smooth curve. The inset shows the individual price intervals.

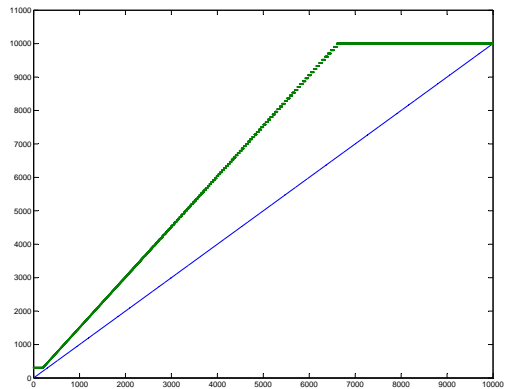

Figure 4: The figure shows the pricing curve drawn by the smoothing algorithm for the linear production curve

### 5.4 The Analysis

Let us call an interval $J_{iq} = [r, s)$ to be *full-sized* if its width equals $\mathsf{width}_i(\pi_i(s))$. Note that the right-most interval $J_{iz_i}$ is not full sized since it semi-infinite. Further, the left-most interval $J_{i1}$ *may not* be full-sized either because the algorithm ran out of copies, or the price became too low so that all remaining unpriced copies were bunched together. We first show that if we sell at least $|J_{i1}| + |J_{i2}|$ copies of item $i$, i.e., we have sold at least one full-sized interval, we get a good approximation factor for the reasons we discussed in Section 5.3. This is proved in Lemma 5.2.

Then we consider the case when the number of items sold is less than $|J_{i1}| + |J_{i2}|$: in this case we cannot show a good multiplicative loss. Instead, we show that the price of items in the first two intervals is small in this case, which bounds the additive loss. This is proved in Lemma 5.5. Finally, our main result Theorem 5.4 follows from these two lemmas.

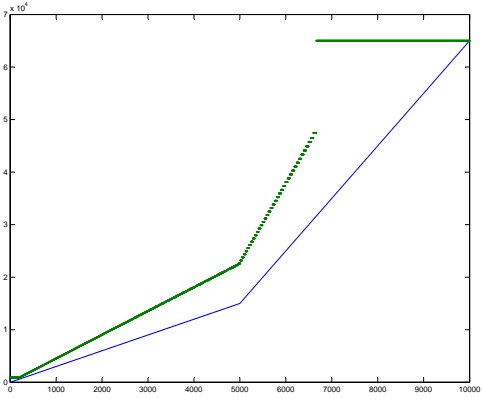
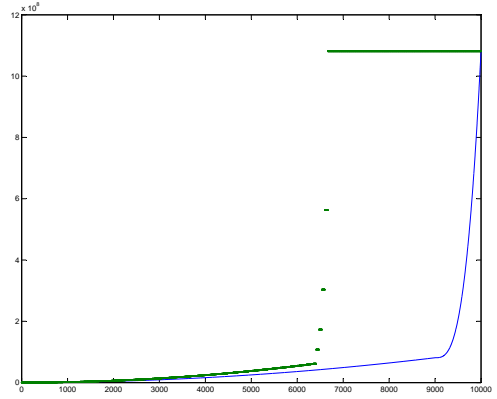

Figure 5: The figure shows the pricing curve drawn by the smoothing algorithm a piece-wise linear production curve. The lower line is the production curve. The upper thicker line is the pricing curve.

Figure 6: The figure shows the pricing curve drawn by the smoothing algorithm a production curve which grows as $x^2$ initially and as $x^3$ in the final phase. The lower line is the production curve. The upper thicker line is the pricing curve.

All the intervals except the leftmost $J_{i1}$ and rightmost $J_{iz_i}$ ones are created in a similar fashion; intervals $J_{i1}$ and $J_{iz_i}$ have to treated as special cases at several points in the analysis. Also, the analysis which follows from this point onwards up till (and not including) Theorem 5.4 is *per item*. Hence the subscript $i$ in the terms involved is irrelevant from the point of analysis and is present only to maintain uniformity in presentation.

To begin, we state some useful properties of the prices and widths of the intervals.

**Lemma 5.1** (Prices and Widths). *The following facts about interval prices hold for the intervals in $\mathcal{J}_i$ for any non-decreasing cost curve:*

  a. *For any $J_{iq} = [r, s)$ such that $q \neq z_i$, $\pi_i(J_{iq}) \geq \frac{3}{2} c_i(s)$. Hence, $\pi_i(x) \geq \frac{3}{2} c_i(x)$ for $x \in J_{iq}$.*
  a'. *If the cost curve is convex, $\pi_i(\lfloor \frac{2}{3} \ell_i \rfloor) \geq \frac{3}{2} c_i(\lfloor \frac{2}{3} \ell_i \rfloor)$.*
  b. *For consecutive $J_{iq}$ and $J_{i\,q+1}$ and $q \neq z_i - 1$, we have $\pi_i(J_{iq}) \leq \pi_i(J_{i\,q+1}) \leq 2\, \pi_i(J_{iq})$. If the cost curve is convex the claim also holds for $q = z_i - 1$.*
  c. *All price intervals $J_{iq} = [r, s)$ ($q \notin \{1, z_i\}$) have $|J_{iq}| = \mathsf{width}_i(\pi_i(s)) = \mathsf{width}_i(\pi_i(J_{i\,q+1}))$.*

Lemma 5.1(a) states that the price of any copy is sufficiently far from the production cost of that copy. Lemma 5.1(a') states the same claim about the left end of the right-most interval $J_{iz_i}$ in case the cost curve is in addition convex. Lemma 5.1(b) states the price of copies in the interval $J_{i\,q+1}$ is higher than that of $J_{i\,q}$, but not too far from it. Lemma 5.1(d) states that all price intervals except possibly the left-most and the right-most are full-sized. Armed with these facts, we first show that if "many" copies of item $i$ are sold, then we are in good shape. The other case where "few" copies are sold, is dealt with subsequently.

### 5.4.1 The Case of Many Copies.

Suppose we sell *all* copies in some interval $J_{iq}$ for $q > 1$: then we get that the profit made from that interval alone gives us a good approximation.

**Lemma 5.2.** *If the number of sold copies $x_i$ of item $i$ is at least $|J_{i1}| + |J_{i2}|$, then $P_i^f \cdot c_i^{invt}(P_i^f) \leq 12 B_i \cdot \mathsf{profit}_i$, where $\mathsf{profit}_i := \sum_{k=1}^{x_i} (\pi_i(k) - c_i(k))$.*

*Proof.* Let $q$ be the largest integer such that $J_{iq} = [r, s)$ is completely sold out; hence $q \in [2, z_i)$. The final price is $P_i^f = \pi_i(J_{i\,q+1}) = \pi_i(s)$. We want to show we make a reasonable profit from the sales of copies in $J_{iq}$. From Lemma 5.1(c), there are $\mathsf{width}_i(\pi_i(s))$ many copies in $J_{iq}$. For each of these copies $k \in [r, s)$, the profit is $\pi_i(k) - c_i(k) \geq \pi_i(k) - c_i(s)$, because costs are non-decreasing.

However, by Step 7 of the pricing algorithm, for all $k \in J_{iq}$, $\pi_i(k) = c_i(s) + \Delta$, where $\Delta$ is determined by Step 6.

- Either $\pi_i(s) \geq 3 c_i(s)$, $\Delta = \frac{1}{2}(\pi_i(s) - c_i(s)) \geq \frac{1}{3}\pi_i(s)$,
- Or $\pi_i(s) < 3 c_i(s)$, $\Delta = c_i(s)/2 > \frac{1}{6}\pi_i(s)$.

So, we make a profit of at least $\pi_i(s)/6$ from each of the $\mathsf{width}_i(\pi_i(s)) = \lfloor \frac{c_i^{invt}(\pi_i(s))}{B_i} \rfloor$ many copies in $J_{iq}$:

$$\mathsf{profit}_i \geq \frac{\pi_i(s)}{6} \cdot \lfloor \frac{c_i^{invt}(\pi_i(s))}{B_i} \rfloor \geq \frac{\pi_i(s) \cdot c_i^{invt}(\pi_i(s))}{12 B_i},$$

where the last inequality is because $\lfloor t \rfloor \geq t/2$ for $t \geq 1$. Plugging in $P_i^f = \pi_i(s)$ completes the proof. $\square$

### 5.4.2 The Case of Few Copies

Now suppose item $i$ is such that the number of copies we sell either lies within the left-most interval $J_{i1}$, or only covers a small fraction of the second interval $J_{i2}$: the argument given above does not hold in that case. However we can show the following result.

**Lemma 5.3.** *If the number of sold copies $x_i$ of item $i$ is less than $|J_{i1}| + |J_{i2}|$ then $\pi_i(P_i^f) \cdot c_i^{invt}(P_i^f) \leq \pi_i(J_{i2}) \cdot c_i^{invt}(\pi_i(J_{i2}))$.*

*Proof.* Since we end up selling less than $|J_{i1}| + |J_{i2}|$ copies, hence the final price $P_i^f$ is at most $\max\{\pi_i(J_{i1}), \pi_i(J_{i2})\}$ which is $\pi_i(J_{i2})$ since Lemma 5.1(b) tell us that $\pi_i(J_{i1}) \leq \pi_i(J_{i2})$. Hence, $P_i^f \cdot c_i^{invt}(P_i^f) \leq \pi_i(J_{i2}) \cdot c_i^{invt}(\pi_i(J_{i2}))$ ($c_i^{invt}(p)$ is non-decreasing function of $p$). $\square$

### 5.4.3 Finishing the Analysis

Lemma 5.2 and Lemma 5.3 together give us the main result of this section.

**Theorem 5.4.** *The social welfare $W(\mathsf{alg})$ achieved by the smoothing algorithm on a non-decreasing cost curve given an estimate $Z \in (U_{max}, U_{max}/\epsilon]$ satisfies*

$$W(\mathsf{alg}) \geq \frac{W(\mathsf{opt}) - \sum_{i \in \mathcal{I}} \pi_i(J_{i2}) \cdot c_i^{invt}(\pi_i(J_{i2}))}{12 \max_{i \in \mathcal{I}} B_i},$$

*where $B_i := \lceil 12 \log(4 n \ell_i / \epsilon) \rceil$, and $\ell_i := \min\{c_i^{inv}(Z), m\}$.*

Roughly, Theorem 5.4 states that the social welfare attained by the smoothing algorithm is a logarithmic approximation to (optimal social welfare minus the *price* of the first few copies of each item).

**Proof of Theorem 5.4 :** For each item $i$, depending on the number of copies sold, either Lemma 5.2 or Lemma 5.3 applies, which implies that for each $i \in \mathcal{I}$,

$$P_i^f \cdot c_i^{\text{invt}}(P_i^f) \leq 12\, B_i \cdot \text{profit}_i + \pi_i(J_{i2}) \cdot c_i^{\text{invt}}(\pi_i(J_{i2})).$$

Summing over all items $i$, we get

$$\sum_{i \in \mathcal{I}} P_i^f \cdot c_i^{\text{invt}}(P_i^f) \leq 12 \max_{i \in \mathcal{I}} B_i \cdot \text{profit}_i + \sum_{i \in \mathcal{I}} \pi_i(J_{i2}) \cdot c_i^{\text{invt}}(\pi_i(J_{i2})).$$

Applying Corollary B.1, we get

$$W(\text{alg}) \geq \frac{W(\text{opt}) - \sum_{i \in \mathcal{I}} \pi_i(J_{i2}) \cdot c_i^{\text{invt}}(\pi_i(J_{i2}))}{12 \max_{i \in \mathcal{I}} B_i},$$

which completes the proof. □

## 5.5 Convex cost curves

Theorem 5.4 leaves us unsatisfied since the additive loss, which is the *price* of the first few copies of each item, is not stated in terms of quantities that are part of the problem statement such as production cost. For convex curves, we are able to overcome that deficiency. The additive loss would be, roughly, the sum of *production cost* of the first few copies of every item. The crucial lemma which we will prove in this section is:

**Lemma 5.5.** *For a convex cost curve, $\pi_i(J_{i2}) \cdot c^{\text{invt}}(\pi_i(J_{i2})) \leq \max\{B_i\, c_i(B_i), \frac{\epsilon Z}{2n}\}$.*

which will suffice to prove the following result.

**Corollary 5.6.** *The social welfare $W(\text{alg})$ achieved by the smoothing algorithm on a non-decreasing convex cost curve given an estimate $Z \in (U_{max}, U_{max}/\epsilon]$ satisfies*

$$W(\text{alg}) \geq \frac{W(\text{opt})/2 - \sum_{i \in \mathcal{I}} B_i \cdot c_i(B_i)}{12 \max_{i \in \mathcal{I}} B_i},$$

*where $B_i := \lceil 12 \log(4n\ell_i/\epsilon) \rceil$, and $\ell_i := \min\{c_i^{\text{inv}}(Z), m\}$.*

Corollary 5.6 gives us the same approximation factor to optimal social welfare as guaranteed by Theorem 5.4, except that it states the additive loss to be the sum of *production cost* of first few copies of each item.

**Proof of Corollary 5.6 :** Putting Theorem 5.4 and Lemma 5.5 together,

$$W(\text{alg}) \geq \frac{W(\text{opt}) - \epsilon Z/2 - \sum_{i \in \mathcal{I}} B_i \cdot c_i(B_i)}{12 \max_{i \in \mathcal{I}} B_i}.$$

Using $\epsilon Z \leq U_{max} \leq W(\text{opt})$, we get the desired result. □

We now need to prove Lemma 5.5. The pricing algorithm terminates when it has priced all the copies, i.e. $x$ is set to 1 and the `if` condition in Step 3 becomes false. $x$ can be set to 1 either in Step 8 (preceded by $x'$ being set to 1 in Step 5) or in Step 11. We consider these two cases separately.

- Algorithm terminates through Step 11: Lemma 5.8 proves that $c_i^{\text{invt}}(\pi_i(J_{i2})) \cdot \pi_i(J_{i2}) < B_i \cdot c_i(B_i)$.

- Algorithm terminates through Step 5: Lemma 5.12 proves that $c_i^{\text{invt}}(\pi_i(J_{i2})) \cdot \pi_i(J_{i2}) < \frac{\epsilon Z}{2n}$.

**Proof of Lemma 5.5 :** The algorithm terminates either through Step 5 or Step 11 and Lemma 5.12 and Lemma 5.8 together indicate that $\pi_i(J_{i2}) \cdot c^{\text{invt}}(\pi_i(J_{i2})) \leq \max\{B_i\, c_i(B_i), \frac{\epsilon Z}{2n}\}$. □

Before proving Lemma 5.8 and Lemma 5.12, we state and prove the following lemma that characterizes the circumstances under which the algorithm terminates in either condition.

**Lemma 5.7.** *The pricing algorithm terminates through Step 11 if and only if $\pi_i(J_{i2}) < c_i(B_i)$.*

*Proof.* Let $J_{i2} = [s, r)$. We first prove that if $\pi_i(J_{i2}) < c_i(B_i)$, then the algorithm terminates in Step 11. If $\pi_i(s) = \pi_i(J_{i2}) < c_i(B_i)$, then it implies that $c_i^{\text{inv}}(\pi_i(s)) < B_i$, and by definition of $c^{\text{invt}}()$, $c_i^{\text{invt}}(\pi_i(s)) < B_i$ which implies that $\text{width}_i(\pi_i(s)) = \lfloor \frac{c_i^{\text{inv}}(\pi_i(s))}{B_i} \rfloor = 0$. Hence, right after creation of $J_{i2}$, when the algorithm checks for the if condition on point $s$ in Step 4, it shall evaluate to false and therefore, the algorithm shall terminate through Step 11.

To prove the other direction, if the algorithm terminates through Step 11, then it must be the case that the if condition in Step 4 evaluated to false for some $x$. Further, $x$ must be the left-end point of $J_{i2}$. This is because once the if condition evaluates to false, the algorithm jumps to Step 11 and creates a single price interval containing all copies that have not been priced yet and it includes the first copy and hence, this price interval must be $J_{i1}$. So $x$ must be the left-end point of the price interval just after $J_{i1}$, i.e. $J_{i2}$.

Now, $\text{width}_i(\pi_i(x)) = \text{width}_i(\pi_i(J_{i2})) = \lfloor \frac{c_i^{\text{invt}}(\pi_i(J_{i2}))}{B_i} \rfloor < 1$ implies that $\frac{c_i^{\text{invt}} \pi_i(J_{i2}))}{B_i} < 1$ and so $c_i^{\text{invt}}(\pi_i(J_{i2})) < B_i$. By definition of $c_i^{\text{invt}}()$, this implies that $\min\{c_i^{\text{inv}}(\pi_i(J_{i2}), \ell_i\} < B_i$. Since by Equation (9), $\ell_i \geq B_i$, it must be the case $c_i^{\text{inv}}(\pi_i(J_{i2})) < B_i$, which by definition of $c_i^{\text{inv}}()$ can occur only if $\pi_i(J_{i2}) < c_i(B_i)$. □

We now prove Lemma 5.8 and Lemma 5.12 that treat the two conditions under which the algorithm can terminate.

**Algorithm terminates through Step 11:** The proof that price of $J_{i2}$ is small follows almost immediately in this case.

**Lemma 5.8.** *If the algorithm terminated through Step 11 then $\pi_i(J_{i2}) \cdot c_i^{\text{invt}}(\pi_i(J_{i2})) < c_i(B_i)\, B_i$.*

*Proof.* If the algorithm terminated through Step 11, then Lemma 5.7 implies that $\pi_i(J_{i2}) < c_i(B_i)$. By definition of $c_i^{\text{invt}}()$, this implies that $c_i^{\text{invt}}(\pi_i(J_{i2})) < B_i$ and hence we get the result. □

**Algorithm terminates through Step 5:** We will prove that price of the interval $J_{i2}$ is 'small' by showing that relative to the price of right-most interval $J_{iz_i}$, the prices for the subsequently created intervals on its left, have been slashed sufficiently often. For item $i$, label a copy $x$ *close* if $\pi_i(x) < 3\, c_i(x)$, else label it as *far*. Depending on which of $r$ and $s$ are close or far, mark a price interval $J_{iq} = [r, s)$ as one of $\{(C, C), (F, C), (C, F), (F, F)\}$. Note that the right-most interval $J_{iz_i}$ is not marked since it is semi-infinite. The following lemma indicates that in case prices are 'far' from the production cost, the algorithm slashes the prices exponentially.

**Lemma 5.9.** *If a contiguous sequence of price intervals $J_{iq}, J_{i\,q+1}, \cdots, J_{i\,q+t-1}$ are all marked $(F,F)$ and $J_{i\,q+t}$ is marked $(F,C)$, then $\pi_i(J_{iq}) \leq (\frac{2}{3})^t \pi_i(J_{i\,q+t})$.*

*Proof.* If interval $J_{ip} = [r,s)$ is marked $(F,F)$ which implies that $3\,c_i(s) < \pi_i(s) = \pi_i(J_{i\,p+1})$, then the pricing algorithm, by Step 6, sets

$$\pi_i(J_{ip}) = c_i(s) + \tfrac{\pi_i(s)-c_i(s)}{2} = \tfrac{\pi_i(s)+c_i(s)}{2} \leq \tfrac{\pi_i(s)+\frac{1}{3}\pi_i(s)}{2} = \tfrac{2}{3}\pi_i(s) = \tfrac{2}{3}\pi_i(J_{i\,p+1}).$$

Hence, $\pi_i(J_{iq}) \leq (\frac{2}{3})\,\pi_i(J_{i\,q+1}) \leq \cdots \leq (\frac{2}{3})^t\,\pi_i(J_{i\,q+t})$. □

Lemma 5.10 states that if we ever have a price interval that is marked $(F,C)$, there are 'many' price intervals to the left of that interval. Lemma 5.11 states that there are 'many' intervals to the left of the right-most interval $J_{iz_i}$.

**Lemma 5.10.** *Consider an interval $J_{iq} = [r,s)$ with $q \neq z_i$ that is marked $(F,C)$. If the algorithm terminated through Step 5, then there are at least $B_i/4$ intervals $J_{iq'}$ with $q' < q$. In particular, $J_{iq}$ cannot be the first price interval i.e. $q \neq 1$.*

*Proof.* Since $s$ is close i.e. $\pi_i(s) < 3\,c_i(s)$, the algorithm, by Step 6, sets $\pi_i(J_{iq}) = c_i(s) + \tfrac{c_i(s)}{2} = \tfrac{3}{2}c_i(s)$. From the definition of $r$ being marked far, $c_i(r) \leq \tfrac{1}{3}\pi_i(r) = \tfrac{1}{3}\tfrac{3}{2}c_i(s) = \tfrac{1}{2}c_i(s)$. Hence, across the interval $J_{iq}$, the cost function increases by at least $\tfrac{1}{2}c_i(s)$. Since $c_i(\cdot)$ is convex, the production cost should rise by at least $\tfrac{1}{2}c_i(s)$ starting from copy $s$ onwards for every $|J_{iq}|$ copies. So, $c_i(r + 5 \cdot |J_{iq}|) = c_i(s + 4 \cdot |J_{iq}|) \leq c_i(s) + 4 \cdot \tfrac{1}{2}(c_i(s)) = 3 \cdot c_i(s)$ and hence

$$c_i^{\mathsf{invt}}(3\,c_i(s)) \leq c_i^{\mathsf{inv}}(3\,c_i(s)) < s + 4 \cdot |J_{iq}| = r + 5 \cdot |J_{iq}|. \tag{10}$$

By Lemma 5.1 (for $q \neq 1$) and Proposition C.5(a) (for $q = 1$), we know that $|J_{iq}| \leq \mathsf{width}_i(\pi_i(s))$. Since $\pi_i(s) < 3\,c_i(s)$, $|J_{iq}| \leq \mathsf{width}_i(\pi_i(s)) \leq \mathsf{width}_i(3\,c_i(s)) \leq \tfrac{c_i^{\mathsf{invt}}(3\,c_i(s))}{B_i}$ where for the second inequality we have used Observation C.1 which says $\mathsf{width}_i(p)$ is a non-decreasing function of $p$. Using (10), we get

$$|J_{iq}| \leq \tfrac{c_i^{\mathsf{invt}}(3\,c_i(s))}{B_i} \leq \tfrac{r + 5 \cdot |J_{iq}|}{B_i} \implies |J_{iq}| \leq \tfrac{r}{B_i - 5}$$

By Equation (9), $B_i \geq 12$ and therefore, $B_i - 5 \geq B_i/2$, hence the above equation implies that $|J_{iq}| \cdot \tfrac{B_i}{2} \leq r$. Since $|J_{iq}| \geq 1$ (any price interval contains at least one copy) and $B \geq 12$, hence $r \geq |J_{iq}| \cdot \tfrac{B_i}{2} \geq 6$. Therefore $q$ cannot be 1, since for $q = 1$, we have $r = 1$ i.e. $J_{i1}$, by definition, is of the form $[1, s)$.

Now, since, $q \neq 1$, $J_{iq}$ cannot contain the first copy i.e. $r \geq 2$ and hence $r - 1 \geq r/2$, and since we already have $|J_{iq}| \cdot \tfrac{B_i}{2} \leq r$, therefore we get,

$$|J_{iq}| \cdot \tfrac{B_i}{4} \leq r - 1 \tag{11}$$

Since the algorithm terminated in Step 5, by Lemma C.6, for all $q' < q$, $|J_{iq'}| \leq \mathsf{width}_i(\pi_i(J_{iq}))$. Moreover, $|J_{iq}| = \mathsf{width}_i(\pi_i(J_{i\,q+1})) \geq \mathsf{width}_i(\pi_i(J_{iq}))$ where the equality follows from Lemma 5.1 and the inequality follows from Observation C.1 and Lemma 5.1(b). Hence, we have that for all $q' < q$, $|J_{iq'}| \leq |J_{iq}|$. Since there are $r - 1$ copies to the left of $J_{iq}$ and for all $q' < q$, $|J_{iq'}| \leq |J_{iq}|$, therefore, by (11), we get the desired result that there are at least $B_i/4$ price intervals $J_{iq'}$ with $q' < q$. □

**Lemma 5.11.** *If the algorithm terminated through Step 5, then there are at least $B_i/3$ intervals $J_{iq}$ with $q < z_i$.*

**Lemma 5.12.** *If the algorithm terminated through Step 5 then $\pi_i(J_{i2}) \cdot c_i^{invt}(\pi_i(J_{i2})) < \frac{\epsilon Z}{2n}$.*

*Proof.* The interval $J_{i1}$ can be marked either $(F,C)$ or $(F,F)$, since $c_i(1) = 0$ while $\pi_i(1) > 0$ (Observation C.2). By Lemma 5.10, $J_{i1}$ cannot be marked $(F,C)$. Hence, the only case left is when $J_{i1}$ is marked $(F,F)$. Let $q$ be the smallest value, if one exists, such that $J_{iq}$ is marked $(F,C)$; note that $q > B_i/4$ by Lemma 5.10, and in particular $q > 2$. If no such $(F,C)$ interval exists, set $q \leftarrow z_i$. By definition of $J_{iq}$, all intervals between $J_{i1}$ and $J_{iq}$ are marked $(F,F)$. Depending on whether $q \neq z_i$ or $q = z_i$, Lemma 5.10 or Lemma 5.11 respectively imply there are at least $B_i/4$ of these intervals. By Lemma 5.9, $\pi_i(J_{i1}) \leq (\frac{2}{3})^{B_i/4} \pi_i(J_{iq}) \leq \frac{\pi_i(J_{iq})}{4n\ell_i/\epsilon}$, since $B_i = \lceil 12 \log(4n\ell_i/\epsilon) \rceil$.

Moreover, by Lemma 5.1(b), $\pi_i(J_{i2}) \leq 2 \cdot \pi_i(J_{i1}) \leq \frac{2\pi_i(J_{iq})}{4n\ell_i/\epsilon}$. By definition of $c^{invt}()$, $c^{invt}(\pi_i(J_{i2})) \leq \ell_i$; this gives $\pi_i(J_{i2}) \cdot c^{invt}(\pi_i(J_{i2})) \leq \frac{2\pi_i(J_{iq})}{4n\ell_i/\epsilon} \cdot \ell_i \leq \frac{\epsilon Z}{2n}$. □

The smoothing algorithm can give purely multiplicative guarantees as long as the cost of the first $O(\log n)$ copies of the items is small compared to $W(\mathsf{opt})$. As an example, suppose the cost functions are $c_i(k) = 0$ for $k \leq d \log n$, and $c_i(k) = \infty$ for $k > d \log n$ for some constant $d$. Then $\ell_i \leq d \log n$, and $B_i = O(\log n/\epsilon)$. So for $d$ large enough constant, $B_i \cdot c(B_i) = 0$, and we get an $O(\log n)$ approximation to the social welfare, as in Bartal et al. (2003). (This is best possible for online algorithms (Awerbuch et al., 1993).)

### 5.5.1 Polynomial production curves

For the case of polynomial production curves of the form[9] $c_i(x) = (x - 1)^d$, we show that the smoothing algorithm gives approximation guarantees close to that of pricing at twice the index; refer Theorem 5.15 and the approximation guarantees given by twice-the-index algorithm on polynomial curves in Section 3.

In the analysis below, we make a few assumptions. First, we assume that $m \geq c_i^{inv}(Z)$ for all items $i$. This case interests us since it is here that the number of copies of an item that are available are less than the number of buyers. In this scenario, for all $p \leq Z$, $c_i^{invt}(p) = c_i^{inv}(p)$ where, recall that $Z$ is the parameter supplied to the smoothing algorithm that satisfies $Z \in (U_{max}, U_{max}/\epsilon]$. Second, we assume that

$$B_i \geq 18(2d+1) \text{ and } \ell_i \geq 2(B_i + 1) \tag{12}$$

These requirements on $\ell_i$ and $B_i$ subsume the ones mentioned in Equation 9.

The crucial result which will help us prove the improved bound is Lemma 5.13.

**Lemma 5.13.** *For all copies $x$ in the range $[B_i, \lfloor (2/3)\ell_i \rfloor - (2d+1) \cdot \lfloor \ell_i/B_i \rfloor]$, $\frac{3}{2} \cdot c_i(x) < \pi_i(x) < 3 \cdot c_i(x)$.*

---

[9]In case the reader is curious on why we choose the polynomial cost curve to be $(x-1)^d$ $(d \geq 1)$ instead of the more natural choice of $x^d$, we recall that the smoothing algorithm analysis assumed that $c_i(1) = 0$ and hence we made the choice of $(x-1)^d$.

The result says that apart possibly from a few copies on the left ($< B_i$) and right ($> \lfloor (2/3)\ell_i \rfloor - (2d+1) \cdot \lfloor \ell_i/B_i \rfloor$) ends, the price is close to the production cost for all copies. We now show how such a result helps in proving in the improved bound. In preparation for applying the structural lemma in Theorem 5.15, the following result gives us the per-item profit equation.

**Lemma 5.14.** *For every item $i$, we have $c_i^{\mathsf{inv}}(P_i^f) \cdot P_i^f \leq 18\,(d+1)\,(27/16)^{d+1}\,\mathsf{profit}_i + 18\,c_i(B_i) \cdot B_i$.*

*Proof.* We consider three cases based on the number of copies $x_i$ of item $i$ that were sold by the smoothing algorithm alg.

- The algorithm sold at most $B_i$ copies of item $i$.

  By Lemma 5.13, the price of $B_i^{th}$ copy is at most $3 \cdot c_i(B_i) = 3\,(B_i - 1)^d$. Since algorithm sold less than $B_i$ copies of item $i$, and by Lemma 5.1(b), prices are non-decreasing from left to right, hence, $P_i^f \leq 3\,(B_i - 1)^d$ and therefore, $c_i^{\mathsf{inv}}(P_i^f) \leq 3^{1/d}(B_i - 1) + 1$. We have $P_i^f \cdot c_i^{\mathsf{inv}}(P_i^f) \leq 3\,(B_i - 1)^d\,(3^{1/d}(B_i - 1) + 1)$.

- The algorithm sold at least $B_i$ copies and less than $\lfloor (2/3)\ell_i \rfloor - (2d+1) \cdot \lfloor \ell_i/B_i \rfloor$ copies.

  Let $x_i$ be the last copy sold. We have $P_i^f = \pi_i(x_i + 1) < 3 \cdot c_i(x_i + 1) = 3 \cdot x_i^d$ where the inequality follows from Lemma 5.13. Hence, $c_i^{\mathsf{inv}}(P_i^f) \leq 3^{1/d} \cdot x_i + 1$. We have $P_i^f \cdot c_i^{\mathsf{inv}}(P_i^f) < (3^{1/d} \cdot x_i + 1) \cdot 3 \cdot (x_i)^d < 6 \cdot 3^{1/d} \cdot x_i^{d+1}$.

  For every copy $x$ up till $x_i$, alg earned profit at least $c_i(x)/2$. From Lemma 5.13, $\pi_i(x) - c_i(x) > c_i(x)/2$. Therefore, the profit $\mathsf{profit}_i$ earned by alg from the sales of item $i$ is at least $\sum_{k=1}^{x_i} \frac{c_i(k)}{2} = \sum_{k=1}^{x_i} \frac{1}{2}(k-1)^d \geq \int_1^{x_i} \frac{1}{2}(k-1)^d \mathrm{d}k \geq \frac{1}{2(d+1)}(x_i - 1)^{d+1}$. Since $x_i \geq B_i$ and by Equation 12, $B_i \geq 18(2d+1) \geq 54$, therefore, $(x_i - 1) \geq (53/54) \cdot x_i$. Therefore, $\mathsf{profit}_i$ is at least $\frac{1}{2(d+1)}(\frac{53}{54})^{d+1} x_i^{d+1}$.

  Hence, we have $c_i^{\mathsf{inv}}(P_i^f) \cdot P_i^f < 2(d+1)(\frac{54}{53})^{d+1} \cdot 6 \cdot 3^{1/d} \cdot \mathsf{profit}_i$.

- The algorithm sold at least $\lfloor (2/3)\ell_i \rfloor - (2d+1) \cdot \lfloor \ell_i/B_i \rfloor$ copies.

  First, we note that $P_i^f \cdot c_i^{\mathsf{inv}}(P_i^f) \leq \ell_i^d \cdot \ell_i = \ell_i^{d+1}$; this follows from the definition of $\ell_i$ and the way we set the price of the right-most price interval. Now following the same argument as in previous case, we know that the profit $\mathsf{profit}_i$ is at least $\sum_{k=1}^{x_i} \frac{1}{2}c_i(x) \geq \frac{1}{2(d+1)}(x_i - 1)^{d+1}$ where $x_i \geq \lfloor (2/3)\ell_i \rfloor - (2d+1) \cdot \lfloor \ell_i/B_i \rfloor$. Since by Equation 12, $B_i \geq 18(2d+1)$, hence, $\lfloor (2/3)\ell_i \rfloor - (2d+1) \cdot \lfloor \ell_i/B_i \rfloor \geq (2/3)\ell_i - 1 - (2d+1) \cdot \ell_i/B_i \geq (2/3)\ell_i - \ell_i/18 - 1 = (11/18)\ell_i - 1$. Hence the profit $\mathsf{profit}_i$ is at least $\frac{1}{2(d+1)}((11/18)\ell_i - 1 - 1)^{d+1} \geq \frac{1}{2(d+1)}(16/27)^{d+1}\ell_i^{d+1}$ where we have used that $((11/18)\ell_i - 2 \geq (16/27)\ell_i$ since by Equation 12, $\ell_i \geq 2(B_i + 1) \geq 2 \cdot (18(2d+1)+1) \geq 110$. Hence $c_i^{\mathsf{inv}}(P_i^f) \cdot P_i^f \leq 2(d+1)(27/16)^{d+1} \cdot \mathsf{profit}_i$.

In all three cases, $P_i^f \cdot c_i^{\mathsf{inv}}(P_i^f) \leq 18\,(d+1)\,(27/16)^{d+1}\mathsf{profit}_i + 3\,(B_i - 1)^d\,(3^{1/d}(B_i - 1) + 1)$. Now note that $c_i(B_i) = (B_i - 1)^d$ and $3\,(3^{1/d}(B_i - 1) + 1) \leq 18\,B_i$. Hence we have the desired result. □

We now present the main result of this section.

**Theorem 5.15.** *The social welfare $W(\mathsf{alg})$ achieved by the smoothing algorithm on the polynomial curve $c_i(x) = (x-1)^d$ satisfies*

$$W(\mathsf{alg}) \geq \frac{W(\mathsf{opt}) - 18 \sum_{i \in \mathcal{I}} B_i \cdot c_i(B_i)}{18\,(d+1)\,(27/16)^{d+1}}$$

where we assume $B_i \geq 18(2d+1)$, $\ell_i \geq 2(B_i+1)$ and that for all items $i$, the number of buyers $m$, exceeds $c_i^{\text{inv}}(Z)$.

Roughly, Theorem 5.15 states that the social welfare attained by the smoothing algorithm on polynomial production curve $(x-1)^d$ is a constant approximation to the optimal social welfare minus the production cost of the first $d$ many copies of each item.

**Proof of Theorem 5.15 :** Summing the result of Lemma 5.14 over all items $i \in \mathcal{I}$, we get

$$\sum_{i \in \mathcal{I}} P_i^f \cdot c_i^{\text{inv}}(P_i^f) \leq 18\,(d+1)\,(27/16)^{d+1}\,\text{profit} + 18 \sum_{i \in \mathcal{I}} c_i(B_i) \cdot B_i$$

Since for all items $i$, $m \geq c_i^{\text{inv}}(Z)$, hence for all $p \leq Z$, $c_i^{\text{invt}}(p) = c_i^{\text{inv}}(p)$. Hence the above result is sufficient for us to apply the structural lemma, Lemma 2.1 and hence we get the desired result. $\square$

We would now like to prove Lemma 5.13. In preparation for that, we shall next prove a few lemmas. The following lemma states for a price interval $J_{iq} = [s,t)$, if at copy $t$, price is close to the production cost, then so it is at copy $s$. This implies in particular that the price is close to production curve for all copies in the price interval $J_{iq}$.

**Lemma 5.16.** *For the polynomial production curve $(x-1)^d$ $(d \geq 1)$, consider a price interval $J_{iq} = [s,t)$ created by the smoothing algorithm such that $t \geq B_i$. If $\pi_i(t) < 3 \cdot c_i(t)$, then $\pi_i(s) < 3 \cdot c_i(s)$. In particular, for all $x \in [s,t)$, $\pi_i(x) < 3 \cdot c_i(x)$.*

*Proof.* First note that since $t \geq B_i$, therefore, $c_i(t) \geq c_i(B_i)$. Also, from Lemma 5.1, we know that $\pi_i(t) \geq \frac{3}{2}c_i(B_i)$. Hence, $c_i^{\text{inv}}(\pi_i(t)) \geq B_i$, and since $c_i^{\text{inv}}() = c_i^{\text{invt}}()$, therefore, it implies that $\text{width}_i(\pi_i(t)) \geq 1$. The width and price of $J_{iq}$ shall therefore be decided by Steps 5- 7.

Since $\pi_i(t) < 3 \cdot c_i(t)$, therefore, $\pi_i(J_{iq}) = \frac{3}{2}c_i(t)$. Further, since $\pi_i(t) < 3 \cdot c_i(t)$, hence $\text{width}_i(\pi_i(t)) = \lfloor c_i^{\text{invt}}(\pi_i(t))/B_i \rfloor \leq \lfloor (3^{1/d} \cdot (t-1) + 1)/B_i \rfloor$ which implies that $s \geq t - 3^{1/d} \cdot (t-1)/B_i - 1$.

Moreover, since $\pi_i(J_{iq}) = \frac{3}{2}c_i(t)$, hence the condition $\pi_i(s) < 3 \cdot c_i(s)$ is equivalent to $\frac{3}{2}c_i(t) < 3 \cdot c_i(s)$ or $c_i(s) > c_i(t)/2$. Since $c_i(x) = (x-1)^d$, therefore, we require $(s-1) > (t-1)/2^{1/d}$. Since $s \geq t - 3^{1/d} \cdot (t-1)/B_i - 1$, it suffices to have $(t-1) - 3^{1/d} \cdot (t-1)/B_i - 1 > (t-1)/2^{1/d}$ which for $t \geq B_i$, is equivalent to demanding $B_i \geq 3^{1/d}/(1 - 1/(t-1) - 1/2^{1/d})$. By Equation 12, we have $t \geq B_i \geq 18(2d+1)$ and hence the inequality is satisfied.

Since for all $x \in [s,t)$, $\pi_i(x) = \pi_i(s)$ and $c_i(x) \geq c_i(s)$, hence, $\pi_i(s) < 3 \cdot c_i(s)$ implies that $\forall x \in [s,t), \pi_i(x) < 3 \cdot c_i(x)$. $\square$

Corollary 5.17 states that in case there is a copy in the range $[B_i, \lfloor (2/3)\,\ell_i \rfloor]$ in the range that is the left end point of a price interval and has its price close to its cost, then for all copies from $B_i$ up till that copy, the price curve is close to the production curve.

**Corollary 5.17.** *If at point $x$ such that $x \in [B_i, \lfloor (2/3)\,\ell_i \rfloor]$, $\pi_i(x) < 3 \cdot c_i(x)$ and $x$ is the left end point of a price interval, then for all copies $x'$ in the range $[B_i, x]$ (i.e. for all copies to the left of $x$ and to the right of $B_i$), $\pi_i(x') < 3 \cdot c_i(x')$.*

*Proof.* Consider a point $x$ such that $\lfloor (2/3)\,\ell_i \rfloor > x > B_i$ and $\pi_i(x) < 3 \cdot c_i(x)$. Say $x$ is the left end point of the price interval $J_{iq}$. Let $J_{i\,q-1} = [r, x)$. By Lemma 5.16, for all copies $y \in J_{i\,q-1}$, $\pi_i(y) < 3\,c_i(y)$. If the left end point $r$ of $J_{i\,q-1}$ is such that $r \leq B_i$, then we have completed the proof of our claim.

Else if the left end point $r$ of $J_{i\,q-1}$ is such that $r > B_i$, we can repeat the argument above since we have $\pi_i(r) < 3\,c_i(r)$ and $r$ is the left end point of $J_{i\,q-1}$ and therefore inductively, we have proved the claim. □

Corollary 5.17 is sufficient to prove Lemma 5.13 in case we can show that a copy which is the left end point of a price interval and has its price close to cost exists in the appropriate range. The following lemma proves the existence of such a copy.

**Lemma 5.18.** *For at least one price interval $J_{ip}$ to the right of the point $\lfloor (2/3)\ell_i \rfloor - (2d+1)\cdot \lfloor \ell_i/B_i \rfloor$, it is the case that the left end point of $J_{ip}$ is marked close.*

*Proof.* Denote the point $\lfloor (2/3)\ell_i \rfloor - (2d+1)\cdot \lfloor \ell_i/B_i \rfloor$ by $w_1$ and the point $\lfloor (2/3)\ell_i \rfloor - 2d\cdot \lfloor \ell_i/B_i \rfloor$ by $w_2$. Let $I$ be the interval $[w_1, w_2]$.

We prove the claim by contradiction; assume that for all price intervals $J_{ip}$ to the right of $w_1$, the left end point of $J_{ip}$ is marked far. Since the width of any price interval is at most $\lfloor \ell_i/B_i \rfloor$ and $|I| = \lfloor \ell_i/B_i \rfloor$, hence there must be a price interval whose left end point, say $\tau$, lies in the interval $I$.

In order to the contradict the assumption, we need to prove that $3\,c_i(\tau) > \pi_i(\tau)$. For this it suffices to show that $3\cdot c_i(w_1) > \pi_i(w_2)$. This is because for any $x \in [w_1, w_2]$, since production curve $c_i()$ is non-decreasing, therefore, $c_i(x) \geq c_i(w_1)$; also by Lemma 5.1(b), $\pi_i(x) \leq \pi_i(w_2)$. Therefore, $3\cdot c_i(w_1) > \pi_i(w_2)$ implies that for any $x$ in $[w_1, w_2]$, $3\cdot c_i(x) \geq 3\cdot c_i(w_1) > \pi_i(w_2) > \pi_i(x)$ and hence in particular $3\,c_i(\tau) > \pi_i(\tau)$.

We now prove that $3\cdot c_i(w_1) > \pi_i(w_2)$. We have

- $c_i(w_1) \geq (\ell_i/2 - (2d+1)\cdot \ell_i/B_i)^d$
  
  This is because $c_i(w_1) = (\lfloor (2/3)\ell_i \rfloor - (2d+1)\cdot \lfloor \ell_i/B_i \rfloor - 1)^d \geq ((2/3)\ell_i - 1 - (2d+1)\cdot \ell_i/B_i - 1)^d \geq (\ell_i/2 - (2d+1)\cdot \ell_i/B_i)^d$ where we use $(2/3)\ell_i - 2 \geq \ell_i/2$ since by Equation 12, $\ell_i \geq 2(B_i+1) \geq 2(18(2d+1)+1) \geq 110$.

- $\pi_i(w_2) = \pi_i(J_{iq}) \leq (\frac{2}{3})^{2d} \cdot \ell_i^d$.

  To see this, let $w_2$ lie in price interval of $J_{iq}$. Note that $J_{iq}$ cannot be the rightmost price interval $J_{iz_i}$ since the left end-point of $J_{iz_i}$ is $\lfloor (2/3)\,\ell_i \rfloor$ and $w_2 < \lfloor (2/3)\ell_i \rfloor$. As all price intervals $J_{iq'}$ to the right of $J_{iq}$ have their left end point marked far, hence, in other words, all price intervals between $J_{iq}$ and $J_{iz_i}$ are marked $(F, F)$. Since a price interval has size at most $\lfloor \ell_i/B_i \rfloor$, therefore, there are at least $2d$ price intervals between $J_{iq}$ and $J_{iz_i}$. By Lemma 5.9, $\pi_i(w_2) = \pi_i(J_{iq}) \leq (\frac{2}{3})^{2d} \cdot \ell_i^d$.

To prove that $3\cdot c_i(w_1) > \pi_i(w_2)$, it suffices to have $(\frac{2}{3})^{2d} \cdot \ell_i^d < 3\cdot (\ell_i/2 - (2d+1)\cdot \ell_i/B_i)^d$ which is equivalent to demanding $(\frac{2}{3})^2 \cdot \frac{1}{3^{1/d}} < \frac{1}{2} - (2d+1)\cdot \frac{1}{B_i}$ or $B_i > (2d+1)/(\frac{1}{2} - (\frac{2}{3})^2 \cdot \frac{1}{3^{1/d}})$; by Equation 12, $B_i > 18(2d+1)$, and hence we have satisfied the desired inequality. □

We now prove Lemma 5.13 which recall, roughly, states that apart from the 'few' left-most copies and right-most copies, the price curve is close to the production curve for all copies.

**Proof of Lemma 5.13 :** From Lemma 5.1(a), we can infer one side of the inequality i.e. for all $x$ in the desired range, $\pi_i(x) \geq 3\,c_i(x)/2$. Now for the other side of the inequality i.e. $\pi_i(x) < 3c_i(x)$.

Denote the point $\lfloor (2/3)\ell_i \rfloor - (2d+1) \cdot \lfloor \ell_i/B_i \rfloor$ by $w$. First we note that $w$ lies to the right of $B_i$ i.e. $w \geq B_i$. In order to see this, it suffices to show $(2/3)\,\ell_i - 1 - (2d+1)\cdot \ell_i/B_i \geq B_i$. By Equation 12, we have $B_i \geq 18(2d+1)$, and hence $(2/3)\ell_i - 1 - (2d+1)\cdot \ell_i/B_i \geq (2/3)\ell_i - 1 - \ell_i/18 = (11/18)\ell_i - 1$. And hence it suffices to have $\ell_i \geq (18/11) \cdot (B_i + 1)$ which by Equation 12 is true.

From Lemma 5.18, we know that there is at least one price interval $J_{ip}$ to the right of $w$ whose left end point say $\tau$ is marked close. Note that $\tau \geq w \geq B_i$. Further since $\tau$ is the *left end point* of a price interval, therefore, $\tau \leq \lfloor (2/3)\ell_i \rfloor$. Hence, we have $B_i \leq \tau \leq \lfloor (2/3)\,\ell_i \rfloor$. Thus, by Corollary 5.17, we get the desired result. $\square$

## 6 Profit Maximization

In this section we show how to combine an online algorithm for social welfare maximization in the presence of increasing costs (such as those in Section 3) with an algorithm for a single-buyer profit maximization (such as the algorithm in Balcan et al. (2008)) to yield an algorithm with strong profit guarantees for any sequence of buyers under increasing costs. Specifically, suppose we are given access to two algorithms:

1. a deterministic social-welfare maximizing algorithm $A$, which given production cost curves $\{c_i\}_{i \in \mathcal{I}}$, outputs pricing schemes $\{\omega_i(\cdot)\}_{i \in \mathcal{I}}$ such that on any sequence $\sigma$ of buyers, the algorithm's social welfare satisfies

$$\rho \cdot W(A(\sigma)) + \beta \geq W(\mathsf{opt}(\sigma)) \tag{13}$$

    We further assume that for every item $i \in \mathcal{I}$ and $k \in \mathbb{N}$, $\omega_i(k) \geq c_i(k)$ i.e. the price of any copy of any item is at least as much as the production cost of that copy of the item.

2. a randomized single-buyer revenue maximization algorithm $B$, which outputs a non-negative price vector $\tau$ for items $i \in \mathcal{I}$ and gives the guarantee that for any buyer $b$, with valuation $v_b()$,

$$\mu \cdot \mathbb{E}_\tau[\sum_{i \in S_\tau} \tau_i] + \kappa \geq \max_{s \subseteq \mathcal{I}} v_b(s) \tag{14}$$

    where $\max_{s \subseteq \mathcal{I}} v_b(s)$, the maximum valuation of the buyer over any set, is an upper bound on maximum profit, and $S_\tau$ is the set of items bought by the buyer $b$ when the price vector $\tau$ is presented so that $\sum_{i \in S_\tau} \tau_i$ is the profit generated from buyer $b$ using price vector $\tau$. Algorithm $B$ operates in a world with zero production costs, and may take as input a parameter $T$ such that $\max_{S \subseteq \mathcal{I}} v_b(S) < T$; the parameters $\mu, \kappa$ may be functions of $T$.

The main result of this section is

**Theorem 6.1.** *Given a $(\rho, \beta)$-social welfare maximization algorithm (as defined in Equation (13)) and a $(\mu, \kappa)$-single buyer profit maximization algorithm (as defined in Equation (14)), we can construct a randomized profit-maximizing algorithm whose expected profit over any sequence $\sigma$ of buyers is at least*

$$\frac{W(\mathsf{opt}(\sigma)) - O(\beta + \kappa \cdot |\sigma|)}{O(\rho + \mu)}\ .$$

We now construct an algorithm $C$ which uses $A$ and $B$, and gives expected profit of approximately $\frac{W(\mathsf{opt}(\sigma))}{(2\rho+8\mu)}$ (with some additive loss) over any sequence $\sigma$ of buyers. The construction of algorithm $C$ and subsequent analysis heavily borrows ideas from a similar result proved in Awerbuch et al. (2003). Our result can be seen as extension of their result to situations with production costs and arbitrary valuations.

**The Algorithm $C$:** In case algorithm $B$ requires an estimate $T$, such that $\max_{S\subseteq\mathcal{I}} v_b(S) < T$, algorithm $C$ takes as input a parameter $T$ such that $U_{max} < T$, where $U_{max}$ is defined as in (5). The parameter $T$ is used each time algorithm $B$ is invoked by $C$.

On a sequence $\sigma$ of buyers, for buyer $j$, let $x_i^j$ denote the number of copies of item $i$ already sold when buyer $j$ walks in.

1. With probability $1/2$, set $t_j = \mathbf{0}$ and with probability $1/2$, generate a random price vector $\tau$ using algorithm $B$ and set $t_j = \tau$.

2. Let the price of each item $i$ be $\omega_i(x_i^j + 1) + t_j(i)$.

Simply put, algorithm $C$ maintain a copy of algorithm $A$ running in the background and keeps updating $A$'s state with the sets buyers are buying. When buyer $j$ walks in, with probability $1/2$, algorithm $C$ presents the price vector as specified by the current state of $A$ (determined by the number of various items sold up till then), and with probability $1/2$, adds a random price vector, generated using $B$, to the price vector specified by $A$.

We now present an analysis of profit generated by algorithm $C$ with the final result mentioned in Theorem 6.1.

**Analysis:** For any allocation $\eta : \mathcal{B} \to 2^{\mathcal{I}}$ (where $\eta(j)$ denotes the set of items allocated to buyer $j$), $\sum_{k<j: i\in\eta_k} 1$ denotes the number of copies of item $i$ allocated to buyer $k < j$ under $\eta$. Hence, $\chi_\eta^j(i) = c_i(1 + \sum_{k<j: i\in\eta(k)} 1)$ denotes the cost of allocating a copy of item $i$ to buyer $j$, given that the buyers previous to her have received their allocations under $\eta$. The cost of allocating $\eta(j)$ to buyer $j$ is therefore $\sum_{i\in\eta(j)} \chi_\eta^j(i)$ and the social welfare achieved by allocating $\eta(j)$ to buyer $j$ is $v_j(\eta(j)) - \sum_{i\in\eta(j)} \chi_\eta^j(i)$.

Let $\hat{s}_j = \mathsf{opt}(j)$ be the set allocated to buyer $j$ under the optimal allocation, $\mathsf{opt}$, when she is part of sequence $\sigma$. Denote by $\hat{\gamma}_j$ the social welfare achieved by allocating $\hat{s}_j$ to $j$, which is equal to $v_j(\hat{s}_j) - \sum_{i\in\hat{s}_j} \chi_{\mathsf{opt}}^j(i)$.

Consider a particular run of algorithm $C$ and for $r < s$, let $t_{r:s}$ denote the set of random choices $t_j$ made for buyers $j \in \{r, r+1, \cdots, s\}$. Let $s_j$, a random variable determined completely by $t_{1:j-1}$, be the set which buyer $j$ would have bought if $t_j$ were chosen to be $\mathbf{0}$, and let $\gamma_j$ be the social welfare achieved by allocating $s_j$ to $j$. Note that $s_j$ *may not* be the set actually bought by buyer $j$, depending on whether or not $t_j$ is zero.

Let $p_j$, a random variable determined completely by $t_{1:j}$, be the profit made by $C$ from buyer $j$. Therefore, the total profit made by algorithm $C$ is $\sum_{j\in Q} p_j$ where $Q$ is the set of buyers.

We partition the set of buyers $Q$ into two sets:

1. Let $Q_1$ be the set of buyers for whom $\gamma_j \geq \frac{1}{2}\hat{\gamma}_j$.

2. Let $Q_2$ be the set of buyer for whom $\gamma_j < \frac{1}{2}\hat{\gamma}_j$.

The partition of $Q$ into $Q_1$ and $Q_2$ is determined by the set of random choices $t_{1:m}$. The following observation follows from the definition of $\hat{\gamma}_j$.

**Observation 6.2.** *The optimal profit is at most the optimal social welfare and that is $\sum_{j \in Q} \hat{\gamma}_j$.*

We now state and prove Lemma 6.3 and Lemma 6.4 which bound the welfare made by the optimal allocation for the buyer sets $Q_1$ and $Q_2$ respectively.

Note that for any $j$, the offsets $t_{1:j-1}$ completely determine the bundles bought by buyers 1 through $j-1$. Given $t_{1:j-1}$, for convenience let us define $\pi_j(s) = \sum_{i \in s} \omega_i(x_i^j + 1)$. $\pi_j(s)$ is equal to the price that would be offered to buyer $j$ for set $s$, in case the offset $t_j$ were chosen to be zero.

**Lemma 6.3** (Low welfare buyers). *For any set of values $t_{1:m}$, $\sum_{j \in Q_2} \hat{\gamma}_j \leq 2\rho \sum_{j \in Q_1 \cup Q_2} p_j + 2\beta$.*

*Proof.* Consider any set of values $t_{1:m}$. Consider a buyer $j$ in $Q_2$. $s_j$ is defined to be the utility-maximizing set if $t_j$ were to be 0, therefore, $v_j(s_j) - \pi_j(s_j) \geq v_j(\hat{s}_j) - \pi_j(\hat{s}_j)$. Now, $\gamma_j \geq v_j(s_j) - \pi_j(s_j)$ and $\hat{\gamma}_j \leq v_j(\hat{s}_j)$. Since buyer $j$ is in $Q_2$, hence $\gamma_j < \frac{1}{2}\hat{\gamma}_j$ and therefore we get

$$v_j(\hat{s}_j) - \pi_j(\hat{s}_j) \leq v_j(s_j) - \pi_j(s_j) \leq \gamma_j < \frac{1}{2}\hat{\gamma}_j \tag{15}$$

which implies that

$$\pi_j(\hat{s}_j) > v_j(\hat{s}_j) - \frac{1}{2}\hat{\gamma}_j \tag{16}$$

Let $s'_j$ be the actual set bought by buyer $j$ from algorithm $C$.

Now consider the sequence $\sigma'$ composed of buyers $j'$ defined as follows

- for every buyer $j$ in $Q_2$, we introduce a buyer $j'$ who has non-zero valuation for exactly two sets – she values set $s'_j$ at $\pi_j(s'_j)$ and set $\hat{s}_j$ at $v_j(\hat{s}_j) - \frac{1}{2}\hat{\gamma}_j$, and

- for every buyer $j$ in $Q_1$, we introduce buyer $j'$, such that she is single-minded and has valuation $\pi_j(s'_j)$ for set $s'_j$.

The sequence of buyers in $\sigma'$ is the natural ordering i.e. $m' < n'$ if and only if $m < n$. It is not difficult to verify that when algorithm $A$ is run on sequence $\sigma'$,

1. for all $j \in Q_1$, buyer $j'$ shall buy the set $s'_j$ from $A$,

2. for all $j \in Q_2$, buyer $j'_1$ shall buy the set $s'_j$ from $A$ (and not the set $\hat{s}_j$ by (16))

Consider the allocation $\eta$ for sequence $\sigma'$, wherein for every $j \in Q_2$, $j'$ is allocated set $\hat{s}_j$ and rest of the buyers are allocated nothing. The social welfare achieved by $\eta$ is $\sum_{j \in Q_2} \left( v_j(\hat{s}_j) - \frac{1}{2}\hat{\gamma}_j - \sum_{i \in \hat{s}_j} \chi_\eta^j(i) \right)$ where for each $j \in Q_2$, $v_j(\hat{s}_j) - \frac{1}{2}\hat{\gamma}_j$ is the value of buyer $j'$ for set $\hat{s}_j$ and $\sum_{i \in \hat{s}_j} \chi_\eta^j(i)$ is the cost of allocating that set. Now observe that for each $j$, $\sum_{i \in \hat{s}_j} \chi_\eta^j(i) \leq \sum_{i \in \hat{s}_j} \chi_{\text{opt}}^j(i)$ i.e. the cost of allocating the set $\hat{s}_j$ to buyer $j'$ under $\eta$ on sequence $\sigma'$ is at most the cost of allocating that set to the buyer $j$ under opt on sequence $\sigma$. This is because for any prefix of buyers, $\eta$ allocates only at most as many copies of any item as opt for that prefix. Therefore, for each $j \in Q_2$, $v_j(\hat{s}_j) - \sum_{i \in \hat{s}_j} \chi_\eta^j(i) \geq v_j(\hat{s}_j) - \sum_{i \in \hat{s}_j} \chi_{\text{opt}}^j(i) = \hat{\gamma}_j$, and therefore, $v_j(\hat{s}_j) - \frac{1}{2}\hat{\gamma}_j - \sum_{i \in \hat{s}_j} \chi_\eta^j(i) \geq \frac{1}{2}\hat{\gamma}_j$. Hence the allocation $\eta$ achieves a social welfare of at least $\sum_{j \in Q_2} \frac{1}{2}\hat{\gamma}_j$

The $(\rho, \beta)$-approximation guarantee of $A$ should hold on $\sigma'$ as well and therefore using (13), and the fact optimal welfare on $\sigma'$ is at least as much the welfare made through allocation $\eta$ we have, $\sum_{j \in Q_2} \frac{1}{2}\hat{\gamma}_j \leq \rho \cdot \sum_{j \in Q_1 \cup Q_2} (\pi_j(s'_j) - c_j(s'_j)) + \beta$.

However, the profit $p_j$ made by $C$ on sequence $\sigma$ is $\pi_j(s'_j) - c_j(s'_j) + \sum_{i \in s'_j} t_j(i)$ and therefore in particular, $p_j \geq \pi_j(s'_j) - c_j(s'_j)$. Hence, we get the desired claim i.e. $\sum_{j \in Q_2} \hat{\gamma}_j \leq 2\rho \sum_{j \in Q_1 \cup Q_2} p_j + 2\beta$. $\square$

**Lemma 6.4** (High welfare buyers). $\mathop{\mathbb{E}}_{t_{1:m}} [\sum_{j \in Q_1} \hat{\gamma}_j] \leq 8\mu \mathop{\mathbb{E}}_{t_{1:m}} [\sum_{j \in Q_1} p_j] + 4\kappa \mathop{\mathbb{E}}_{t_{1:m}} [|Q_1|]$.

*Proof.* For a buyer $j$ in $Q_1$, we know that $\gamma_j = (v_j(s_j) - c_j(s_j)) \geq \frac{1}{2}\hat{\gamma}_j$.

- Either, $(\pi_j(s_j) - c_j(s_j)) \geq \frac{1}{2}\gamma_j \geq \frac{1}{4}\hat{\gamma}_j$: With probability $1/2$, we choose $t_j = \mathbf{0}$, and by definition of $s_j$, we know that buyer $j$ would buy set $s_j$ and therefore the profit from buyer $j$, $p_j = (\pi_j(s_j) - c_j(s_j)) \geq \frac{1}{2}\gamma_j \geq \frac{1}{4}\hat{\gamma}_j$.

- Or, $(\pi_j(s_j) - c_j(s_j)) < \frac{1}{2}\gamma_j$. In this case, $v_j(s_j) - \pi_j(s_j) \geq \frac{1}{2}\gamma_j$ because $(v_j(s_j) - \pi_j(s_j)) + (\pi_j(s_j) - c_j(s_j)) = \gamma_j$.

  With probability $1/2$, we set $t_j$ to be a random vector $\tau$ generated using algorithm B. Consider a setting with zero production cost and a buyer $b$ whose valuation $v_b()$ is given as $\forall s \subseteq \mathcal{I}, v_b(s) = v_j(s) - \pi_j(s)$. For any $\tau$ and for any set $s$, buyer $j$ and buyer $b$ have the same utility as we can see in the following equation:

  $$v_b(s) - \sum_{i \in s} \tau_i = v_j(s) - \pi_j(s) - \sum_{i \in s} \tau_i$$

  Hence on being presented with price vector $\tau$, buyer $b$ shall the buy the same set as buyer $j$, call the set $S_\tau$. Therefore, the expected value of $\sum_{i \in S_\tau} \tau_i$ is equal to the expected profit made from buyer $b$ which by (14) is

  $$\frac{\max_{s \subseteq \mathcal{I}} v_b(s) - \kappa}{\mu}.$$

  Since we are in the case where $(\pi_j(s_j) - c_j(s_j)) < \frac{1}{2}\gamma_j$, therefore $\max_{s \subseteq \mathcal{I}} v_b(s) = v_j(s_j) - \pi_j(s_j) \geq \frac{1}{2}\gamma_j$. Since for buyer $j$, we choose a random offset $\tau$ with probability $1/2$, therefore, the expected profit from buyer $j$ is at least

  $$\frac{\max_{s \subseteq \mathcal{I}} v_b(s) - \kappa}{2\mu} \geq \frac{\frac{1}{2}\gamma_j - \kappa}{2\mu}.$$

Therefore, taking both of the above cases into account, for any buyer $j \in Q_1$,

$$\mathop{\mathbb{E}}_{t_j | t_{1:j-1}} [p_j] \geq \frac{\frac{1}{2}\gamma_j - \kappa}{2\mu} \geq \frac{\frac{1}{4}\hat{\gamma}_j - \kappa}{2\mu}.$$

Taking expectation over $t_{1:j}$, we get

$$\underset{t_{1:j}}{\mathbb{E}}[p_j \cdot \mathbb{I}[j \in Q_1]] \geq \frac{\underset{t_{1:j}}{\mathbb{E}}[(\frac{1}{4}\hat{\gamma}_j - \kappa) \cdot \mathbb{I}[j \in Q_1]]}{2\mu}$$

where $\mathbb{I}[\cdot]$ is the indicator function. The expectation can be extended to be over $t_{1:m}$ since $t_{1:j}$ completely determine $p_j$ and $\mathbb{I}[j \in Q_1]$. Therefore,

$$\underset{t_{1:m}}{\mathbb{E}}[p_j \cdot \mathbb{I}[j \in Q_1]] \geq \frac{\underset{t_{1:m}}{\mathbb{E}}[(\frac{1}{4}\hat{\gamma}_j - \kappa) \cdot \mathbb{I}[j \in Q_1]]}{2\mu}$$

Using Linearity of expectation, we get the desired result. □

We now state the main theorem which as we later show is equivalent to Theorem 6.1 and thereby prove the claimed profit guarantee.

**Theorem 6.5** (Profit Guarantee).

$$\sum_{j \in Q} \hat{\gamma}_j \leq (2\rho + 8\mu) \mathbb{E}[\sum_{j \in Q} p_j] + 4\kappa |Q| + 2\beta.$$

*Proof.* From Lemma 6.3, we get that for any set of values $t_{1:m}$, $\sum_{j \in Q_2} \hat{\gamma}_j \leq 2\rho \sum_{j \in Q_1 \cup Q_2} p_j + 2\beta$ and therefore, $\underset{t_{1:m}}{\mathbb{E}}[\sum_{j \in Q_2} \hat{\gamma}_j] \leq 2\rho \underset{t_{1:m}}{\mathbb{E}}[\sum_{j \in Q} p_j] + 2\beta$.

From Lemma 6.4, we get

$$\underset{t_{1:m}}{\mathbb{E}}[\sum_{j \in Q_1} \hat{\gamma}_j] \leq 8\mu \underset{t_{1:m}}{\mathbb{E}}[\sum_{j \in Q_1} p_j] + 4\kappa \mathbb{E}[|Q_1|].$$

Hence using Linearity of Expectation, we get

$$\sum_{j \in Q} \hat{\gamma}_j \leq (2\rho + 8\mu) \underset{t_{1:m}}{\mathbb{E}}[\sum_{j \in Q} p_j] + 4\kappa |Q| + 2\beta.$$

□

**Proof of Theorem 6.1 :** Recall that the optimal social welfare, and hence the optimal profit, on sequence $\sigma$ is upper-bounded by $\sum_{j \in Q} \hat{\gamma}_j$ (Observation 6.2), while the expected profit generated by algorithm is $C$ is given by $\mathbb{E}[\sum_{j \in Q} p_j]$. Hence Theorem 6.5 is equivalent to result quoted in Theorem 6.1. □

**Remark 6.6.** 1. *In case the social-welfare maximizing algorithm A takes estimate of $U_{max}$: Suppose the estimate given to algorithm C (which passes it on to the copy of A running in the background) is that $U_{max} \in [\delta Z, Z)$. Note that the only place where we use the guarantee is in Lemma 6.3. In the proof, in the stream $\sigma'$, add a fake $\delta Z$- valuation buyer at the end of the stream to make the guarantee hold. The profit guarantee changes to*

$$\sum_{j \in Q} \hat{\gamma}_j \leq (2\rho + 8\mu) \mathbb{E}[\sum_{j \in Q} p_j] + 4\kappa |Q| + 2\beta + 2\rho \delta Z$$

2. *Balcan et al. (2008)* give a single-buyer profit maximization algorithm under zero production cost, which with slight modification, given a parameter $T > U_{max}$, has values of parameters $\kappa = \frac{T}{2\,m\,n}$ and $\mu = O(\log(m\,n))$. This profit maximization algorithm picks a uniform price on a geometric scale for all items and can be combined with either of the social welfare maximizing algorithms in this paper to give a $O(\log(m\,n))$-profit maximizing algorithm with some additive loss.

# A Some Illustrative Examples

## A.1 Some 'natural' pricing schemes

We give some natural pricing schemes and instances where they fail to achieve good social welfare.

### A.1.1 Pricing at Cost

While the algorithm of pricing *at cost* (i.e., setting $\pi(k) = c(k)$) gives an optimal welfare for the unlimited supply setting (where production costs are zero), it is not a good algorithm even for "simple" cost curves. E.g., for a single item with linear costs $c(k) = k$, consider a sequence of $m$ buyers with the $i^{th}$ buyer having value $i$ for $i \in \{1, \ldots, m\}$, followed by $m$ buyers with value $m$ each. Pricing at cost will sell to the first $m$ buyers and give zero welfare for them, after which the production cost will be too high to sell any further copies. In contrast, the optimal solution is to sell to the second set of $m$ buyers with welfare $m^2 - \frac{m(m+1)}{2} = \Omega(m^2)$.

### A.1.2 Pricing at Twice the Cost

Another natural algorithm is to price at twice (or any fixed multiple) of the *cost* of each item. However, while this can be shown to perform well for linear and low-degree polynomial cost functions, it performs poorly for the case of logarithmic costs. Indeed, consider a single item with production cost $c(x) = \log x$, and suppose we price the $i^{th}$ item at cost $\pi(i) = 2 \log i$. Suppose the first $m$ buyers have valuations $2 \log 1, 2 \log 2, \ldots, 2 \log m$ respectively, and are followed by $m^2$ buyers with valuation $2 \log m = \log m^2$. The algorithm would sell to the first $m$ buyers, getting a social welfare of $\sum_{i=1}^{m}(2 \log i - \log i) = O(m \log m)$, after which the cost would be too high for the remaining buyers. In contrast, optimum would sell to the last $m^2$ buyers, and get a social welfare of $\sum_{i=1}^{m^2}(\log m^2 - \log i) = \Omega(m^2)$.

## A.2 Pricing at Twice the Index

Here is an example where twice-the-index algorithm fails to produce good social welfare—e.g., consider the limited supply-like setting where $c(k) = 0$ for $k \leq B$, and $c(k) = V$ for $k > B$. Consider sending in $B$ buyers with valuation zero, followed by $B$ buyers with valuation $V - \varepsilon$. Twice-the-index prices the first $B/2$ copies at zero, and the rest at $V$, whence we get zero welfare, whereas the optimal welfare of $B(V - \varepsilon)$ is achieved by selling to just the later $B$ buyers.

## A.3 The Necessity of Additive Loss

If we do not have an estimates for $W(\mathsf{opt})$, we give a trade-off between the additive and multiplicative loss (even for a single item), for any algorithm where the prices are at least the production cost.

**Lemma A.1.** *With no estimate of $W(\mathsf{opt})$ it is impossible for a deterministic algorithm to give a purely multiplicative guarantee i.e. a guarantee of the form*

$$W(\mathsf{alg}) \geq W(\mathsf{opt})/\alpha$$

*for any finite $\alpha$.*

*Proof.* Suppose we have an algorithm $\mathcal{A}$ that gets such a $\alpha$-approximation for all inputs. Consider a single item with production cost function $c(k) = k$. Suppose the price of the first copy is set to any $1 + \theta$, for $\theta > 0$. Then we can send in a single buyer with valuation $1 + \theta - \varepsilon$, getting a zero social welfare, whereas the optimal welfare is $\theta - \varepsilon > 0$. On the other hand, if the price of the first copy is 1, then first send a buyer with value 1, and then a buyer with value 1.9—the optimal welfare of 0.9 is achieved by selling to the second buyer, but we only sell to the first buyer, get zero welfare again. □

### A.3.1 Some Quantitative Trade-offs

**Lemma A.2.** *For any deterministic pricing algorithm (in a single item setting) acting on production costs $c()$ and that price copies at at least their production cost, to give the guarantee $W(\mathsf{alg}) \geq (W(\mathsf{opt}) - \Delta)/\alpha$, it is necessary that $\alpha \geq \frac{c(2)-c(1)}{\Delta} - 1$.*

*Proof.* Let $\pi(1) = c(1) + \gamma$. Note that $\gamma \leq \Delta$ because otherwise a buyer sent in with valuation $c(1) + \gamma - \varepsilon$ would buy nothing and hence $W(\mathsf{alg}) = 0$ while $W(\mathsf{opt}) = \gamma - \epsilon$ and therefore $W(\mathsf{alg}) \geq (W(\mathsf{opt}) - \Delta)/\alpha$ would be false.

Now consider a sequence of two buyers, the first with valuation $c(1) + \gamma$ and the second with valuation $c(2) - \varepsilon$. The first buyer will buy the first copy. Since the price of second copy is at least $c(2)$, hence the second buyer won't buy. Hence, $W(\mathsf{alg}) = \gamma$ while $W(\mathsf{opt}) = c(2) - c(1) - \varepsilon$. In such a scenario, for the guarantee to hold we require that $\gamma \geq (c(2) - c(1) - \varepsilon - \Delta)/\alpha$ which implies that $\gamma \alpha + \Delta \geq c(2) - c(1) - \varepsilon$. Noting that $\gamma \leq \Delta$ and that the inequality needs to hold for any $\varepsilon \geq 0$, the claim follows. □

The following corollary follows immediately.

**Corollary A.3.** *For production curves $c(x) = x^d$, for $\alpha = 4d$, $\Delta = \Omega(2^d/d)$.*

## B  Variant of Structural Lemma

We now prove a variant of the structural theorem. Define $c_i^{\mathsf{invt}}(p) = \min\{c_i^{\mathsf{inv}}(p), m, c_i^{\mathsf{inv}}(U_{max})\}$ where $m$ is the number of buyers and for a given set of buyers $\mathcal{B}$ and items $\mathcal{I}$,

$$U_{max} = \max_{b \in \mathcal{B}} \max_{T \subseteq \mathcal{I}} \left( v_b(T) - \sum_{i \in T} c_i(1) \right)$$

is the maximum welfare any single buyer can achieve.

**Corollary B.1.** *For a pricing algorithm $\mathsf{alg}$ with non-decreasing price functions $\pi_i$ suppose there exists some $\alpha \geq 1$ and $\beta \geq 0$ such that for every allowed set of values of the final prices $P_i^f$,*

$$\sum_{i \in \mathcal{I}} \sum_{k=1}^{c_i^{\mathsf{invt}}(P_i^f)} (P_i^f - c_i(k)) \leq \alpha \sum_{i \in \mathcal{I}} \mathsf{profit}_i + \beta , \tag{17}$$

*then on every instance of buyers $W(\mathsf{alg}) \geq \frac{1}{\alpha}(W(\mathsf{opt}) - \beta)$.*

**Proof Sketch:** Note that in the proof of Lemma 2.1 just after Equation (3), we argued that $\lambda_i \leq c_i^{\mathsf{inv}}(P_i^f)$. Instead of summing all the way to $c_i^{\mathsf{inv}}(P_i^f)$, we could stop the summation at $\min\{c_i^{\mathsf{inv}}(P_i^f), m, c_i^{\mathsf{inv}}(U_{max})\}$. Indeed, this is because

- $\lambda_i \leq m$: each buyer wants at most one copy of each item, so at most $m$ copies of item $i$ can be allocated in the optimal solution.

- $\lambda_i \leq c_i^{\mathsf{inv}}(U_{max})$: each copy beyond $c_i^{\mathsf{inv}}(U_{max})$ has cost strictly greater than $U_{max}$; allocation of any such copy can only decrease the social welfare.

□

## C Some observations and results for Section 5

We now state two observations which are easy to prove.

**Observation C.1.** $\mathsf{width}_i(p)$ *is non-decreasing in* $p$.

**Observation C.2.** *Assuming the parameter $Z > 0$, for every copy $x$, the price set by the algorithm, $\pi_i(x) > 0$.*

**Claim C.3.** *In the analysis of the smoothing algorithm, it is sufficient to consider only those items that have $\ell_i \geq B_i$.*

**Proof Sketch:** We would like to show that we can assume $\ell_i \geq B_i \geq 12$ without loss of generality. We first show that $B_i \geq 12$. Recall that $\ell_i := \min\{c_i^{\mathsf{inv}}(Z), m\}$ and $B_i = \lceil 12 \log(4n\ell_i/\epsilon) \rceil$. We can assume that $U_{max} > 0$, so $Z > 0$, and since $c_i(1) = 0$, hence $\ell_i \geq 1$; in turn this implies that $B_i \geq 12$.

Now, if the minimum $\ell_i < B_i$ because $m$ is small, we can always add in dummy buyers, this does not change any of the arguments. Else, it must be the case that $\ell_i = c_i^{\mathsf{inv}}(Z) < B_i$, which means $c_i(B_i) > Z \geq U_{max}$. We claim that we can just drop all such items from the instance, and run our algorithm on the remaining items, with guarantees identical to those in Theorem 5.4.

Indeed, how many copies of item $i$ could we possibly sell in the optimal solution? At most $\ell_i$, since after that its cost is at least $c_i(\ell_i) \geq Z$, too high for opt to allocate to anyone without decreasing the social welfare as the cost exceeds the valuation. Therefore, since at most $\ell_i$ copies of such an item can be allocated, so ignoring this item entirely can drop $W(\mathsf{opt})$ by at most $\ell_i \cdot U_{max} < B_i \cdot c_i(B_i)$. Hence, dropping all such items implies that the remaining set of items $\mathcal{I}'$ (and the original set of buyers) have an modified optimal welfare of $W(\mathsf{opt}') \geq W(\mathsf{opt}) - \sum_{i \in \mathcal{I}\setminus\mathcal{I}'} B_i \cdot c_i(B_i)$. For this new instance, Theorem 5.4 gives a welfare of

$$W(\mathsf{alg}) \geq \frac{W(\mathsf{opt}')/2 - \sum_{i \in \mathcal{I}'} B_i \cdot c_i(B_i)}{12 \max_{i \in \mathcal{I}'} B_i} \geq \frac{W(\mathsf{opt})/2 - \sum_{i \in \mathcal{I}} B_i \cdot c_i(B_i)}{12 \max_{i \in \mathcal{I}} B_i} \tag{18}$$

Hence, we can assume $\ell_i \geq B_i \geq 12$ without loss of generality. □

**Lemma C.4.** *The number of price intervals, $z_i \geq 3$*

*Proof.* $z_i \neq 1$ since the first time the algorithm checks for condition $x > 1$ in Step 3, it evaluates to true because $x$ is set to $\lfloor \frac{2}{3}\ell_i \rfloor$ by Step 2 and since by Equation (9), $\ell_i > 3$, therefore $x = \lfloor \frac{2}{3}\ell_i \rfloor \geq \ell_i/3 > 1$. Hence, the algorithm creates at least one price interval other than $[\lfloor \frac{2}{3}\ell_i \rfloor, \infty)$.

We now prove that $z_i$ is at least 3. We prove by contradiction. If it were the case $z_i = 2$, then it implies that the algorithm terminates the second time it checks for the condition $x > 1$ in Step 3.

As observed earlier, $x$ can be set to 1 either by Step 5 or by Step 11. To disambiguate let the value of $x$ be $x_1$ and $x_2$ the first and second time respectively, the `while` loop condition at Step 3 is checked. We know from Step 2, that $x_1 = \lfloor \frac{2}{3} \ell_i \rfloor$. For $z_i$ to be 2, we require $x_2$ to be 1.

- If $x_2$ is set to 1 by Step 11, it implies that the condition $\mathsf{width}_i(\pi_i(x_1)) \geq 1$ in Step 4 must have evaluated to false. However, $\pi_i(x_1) = Z$ (by Step 1) and therefore, $\mathsf{width}_i(\pi_i(x_1)) = \lfloor c_i^{\mathsf{invt}}(Z)/B_i \rfloor$. Now $c_i^{\mathsf{invt}}(Z) = \min\{c_i^{\mathsf{inv}}(Z), \ell_i\}$ and $\ell_i = \min\{c_i^{\mathsf{inv}}(Z), m\}$ and therefore, $c_i^{\mathsf{invt}}(Z) = \ell_i$. Hence, $\mathsf{width}_i(\pi_i(x_1)) = \lfloor \ell_i/B_i \rfloor \geq 1$ since $\ell_i \geq B_i$ by Equation (9). Hence, $x_2$ could not have been set to 1 by Step 11.

- The other case is that $x_2$ is set to 1 by Step 5. This implies that $\max\{x_1 - \mathsf{width}_i(\pi_i(x_1)), 1\} = 1$. However, $x_1 - \mathsf{width}_i(\pi_i(x_1)) = \lfloor \frac{2}{3} \ell_i \rfloor - \lfloor \ell_i/B_i \rfloor \geq \ell_i/3 - \ell_i/B_i \geq 2$ which is satisfied due to Equation (9). Therefore, $x_2 > 2$ and hence could not have been set to 1 by Step 5.

This proves the contradiction. $\square$

**Proposition C.5** (The left-most interval). *The following facts hold for the left-most interval $J_{i1}$:*

a. *If the procedure terminated through Step 5 creating $J_{i1} = [1, s)$, then $|J_{i1}| \leq \mathsf{width}_i(\pi_i(s)) = \mathsf{width}_i(\pi_i(J_{i2}))$.*

b. *If the procedure terminated through Step 11, then $\pi_i(J_{i1}) = \pi_i(J_{i2})$.*

*Proof.* If the algorithm terminated through Step 5, then by construction we have $|J_{i1}| \leq \mathsf{width}_i(\pi_i(s)) = \mathsf{width}_i(\pi_i(J_{i2}))$. If the algorithm terminated through Step 11, then we have no non-trivial bound on $|J_{i1}|$, however, by Step 10, we have $\pi_i(J_{i1}) = \pi_i(J_{i2})$. $\square$

**Proof of Lemma 5.1 :**

- Part (a): We first prove that for $J_{iq} = [r, s)$, $\pi_i(J_{iq}) > \frac{3}{2} c_i(s)$. First consider the case $q \neq 1$.

  - either $\pi_i(s) \geq 3\, c_i(s)$, in which case,
  
  $$\pi_i(J_{iq}) = c_i(s) + \frac{\pi_i(s) - c_i(s)}{2} = \frac{\pi_i(s) + c_i(s)}{2} \geq \frac{3\, c_i(s) + c_i(s)}{2} = 2\, c_i(s).$$
  
  - or, $\pi_i(s) < 3\, c_i(s)$, in which case,
  
  $$\pi_i(J_{iq}) = c_i(s) + \frac{c_i(s)}{2} = \frac{3}{2} c_i(s).$$

  In both cases, the inequality $\pi_i(J_{iq}) \geq \frac{3}{2} c_i(s)$ is true. Now for the case $q = 1$: the above argument also holds if the algorithm terminated in Step 5. If however the algorithm terminated in Step 11, then let $J_{i1} = [1, r)$ and $J_{i2} = [r, s)$ ($J_{i2} \neq J_{iz_i}$ by Lemma C.4). The observations

  1. $\pi_i(J_{i1}) = \pi_i(J_{i2})$ implied by Proposition C.5(b),
  2. $\pi_i(J_{i2}) \geq \frac{3}{2} c_i(s)$, which is at least $\frac{3}{2} c_i(r)$, the first implied by the above argument for $q \neq 1$ and the second implied by monotonicity of $c_i()$.

  together imply the result for $J_{i1}$.

  Having proved that $\pi_i(J_{iq}) > \frac{3}{2} c_i(s)$, note that since $c_i()$ is non-decreasing, therefore we have $c_i(x) \leq c_i(s)$ for all $x \in [r, s)$, so $\pi_i(x) = \pi_i(J_{iq}) \geq \frac{3}{2} c_i(s) \geq \frac{3}{2} c_i(x)$ which proves the second part of the claim.

- For part (a'), convexity implies that $c_i(\lfloor \frac{2}{3} \ell_i \rfloor) \leq \frac{2}{3} c_i(\ell_i)$. By the definition of $\ell_i$, this is at most $\frac{2}{3} Z$. On the other hand, $\pi_i(\lfloor \frac{2}{3} \ell_i \rfloor) = \pi_i(J_{iz_i}) = Z$. Therefore, $\pi_i(\lfloor \frac{2}{3} \ell_i \rfloor) \geq \frac{3}{2} c_i(\lfloor \frac{2}{3} \ell_i \rfloor)$.

- For part (b), first consider the case where $q \notin \{1, z_i - 1\}$, where $J_{iq} = [r, s)$ and $J_{i\,q+1} = [s, t)$.

  - Either $\pi_i(s) \geq 3 c_i(s)$: then

  $$\pi_i(J_{iq}) = c_i(s) + \frac{\pi_i(s) - c_i(s)}{2} = \frac{\pi_i(s) + c_i(s)}{2} \leq \frac{\pi_i(s) + \frac{1}{3}\pi_i(s)}{2} = \frac{2}{3}\pi_i(s) = \frac{2}{3}\pi_i(J_{i\,q+1}).$$

  Also, $\pi_i(J_{iq}) = \frac{\pi_i(s)+c_i(s)}{2} \geq \frac{\pi_i(s)}{2} = \frac{1}{2}\pi_i(J_{i\,q+1})$. Hence, $\pi_i(J_{iq}) \leq \pi_i(J_{i\,q+1}) \leq 2\pi_i(J_{iq})$.

  - Or $\pi_i(s) < 3 c_i(s)$: then

  $$\pi_i(J_{iq}) = c_i(s) + \frac{c_i(s)}{2} = \frac{3}{2} c_i(s) \leq \frac{3}{2} c_i(t) \leq \pi_i(t) = \pi_i(J_{i\,q+1})$$

  where the first inequality follows from the monotonicity of $c_i$, and the second from Lemma 5.1(a). Further, $\pi_i(J_{iq}) = \frac{3}{2} c_i(s) > \frac{1}{2}\pi_i(s) = \frac{1}{2}\pi_i(J_{i\,q+1})$. Hence, we get $\pi_i(J_{iq}) \leq \pi_i(J_{i\,q+1}) \leq 2\pi_i(J_{iq})$.

  Now for the case of $J_{i1}$ ($q = 1$). Note that from Lemma C.4, $z_i \geq 3$. Therefore in particular, $J_{i2} \neq J_{iz_i}$. The analysis above for $J_{iq}$ also holds for $J_{i1}$ if the algorithm terminated in Step 5. Otherwise, by Proposition C.5(b), $\pi_i(J_{i1}) = \pi_i(J_{i2})$ in which case the both the inequalities trivially follow.

  Finally for the case of $q = z_i - 1$. Let $J_{i\,z_i-1} = [r, s)$ and $J_{i\,z_i} = [s, \infty)$. Note that since $s$ is left end point of $J_{iz_i}$, hence $s = \lfloor \frac{2}{3} \ell_i \rfloor$. Either $\pi_i(s) \geq 3 c_i(s)$ in which case the analysis above for $q \notin \{1, z_i - 1\}$ holds for $q = z_i - 1$ as well and shows that $\pi_i(J_{i\,z_i-1}) \leq \pi_i(J_{iz_i}) \leq 2\pi_i(J_{i\,z_i-1})$; or $\pi_i(s) < 3 c_i(s)$, in this case

  $$\pi_i(J_{i\,z_i-1}) = c_i(s) + \frac{c_i(s)}{2} = \frac{3}{2} c_i(s) \leq \pi_i(s) = \pi_i(J_{iz_i})$$

  where the second inequality follows from Lemma 5.1(a').

- Part (c): By construction (Step 5-7), for all price intervals $J_{iq} = [r, s)$ (except maybe $J_{i1}$ and $J_{iz_i}$) we have $|J_{iq}| = \text{width}_i(\pi_i(s))$. Since $s \in J_{i\,q+1}$, therefore, $\pi_i(s) = \pi_i(J_{i\,q+1})$ and hence we have $|J_{iq}| = \text{width}_i(\pi_i(s)) = \text{width}_i(\pi_i(J_{i\,q+1}))$.

□

## C.1 Proofs from Section 5.4.2 for convex cost curves

**Lemma C.6.** *If the pricing algorithm terminated through Step 5, then for any $J_{iq}$, it is true that for all $q' < q$, $|J_{iq'}| < \text{width}_i(\pi_i(J_{iq}))$.*

*Proof.* Consider any $J_{iq}$. By Lemma 5.1(b), we know that for $q' < q$, $\pi_i(J_{iq'}) \leq \pi_i(J_{i\,q'+1}) \leq \pi_i(J_{iq})$. Hence, for all price intervals $q' < q$ and $q' \neq 1$, by Lemma 5.1(c), $|J_{iq'}| = \text{width}_i(\pi_i(J_{i\,q'+1})) \leq \text{width}_i(\pi_i(J_{iq}))$ where the last inequality follows by Observation C.1. Further, by Proposition C.5(a), $|J_{i1}| \leq \text{width}_i(\pi_i(J_{i2})) \leq \text{width}_i(\pi_i(J_{iq}))$. This finishes the proof. □

**Proof of Lemma 5.11:** Note that $\pi_i(J_{iz_i}) = Z$ and so $c_i^{\text{invt}}(\pi_i(J_{iz_i})) = \ell_i$. Consequently, $\text{width}_i(\pi_i(J_{iz_i})) \leq \frac{\ell_i}{B_i}$. Since the algorithm terminated through Step 5, hence by Lemma C.6, for all $q' < z_i$, $|J_{iq'}| \leq \text{width}_i(\pi_i(J_{iz_i})) \leq \frac{\ell_i}{B_i}$.

Since we have $\lfloor 2 \cdot \ell_i/3 \rfloor - 1$ copies to the left of $J_{iz_i}$ (which due to Equation (9) is at least $\ell_i/3$), therefore, the number of intervals $J_{iq}$ with $q < z_i$ is least $\frac{\ell_i/3}{\ell_i/B_i} = B_i/3$. $\square$

## D  Translating the Cost Curve

We show that it is fine to translate the cost functions $c_i()$ to satisfy $c_i(1) = 0$.

**Lemma D.1.** *Given a pricing algorithm $\mathcal{A}'$ for production cost curves $\{c_i'()\}$, which for any set of buyers achieves a guarantee of $(W(\text{opt}) - \beta)/\alpha$, we can create a pricing algorithm $\mathcal{A}$ for the cost curves $c_i(x) = c_i'(x) + \delta_i \mathbf{1}_{x>0}$ for constants $\delta_i \geq 0$, that achieves the same guarantees.*

*Moreover, in case $\mathcal{A}'$ needs an estimate of $\max_{b \in \mathcal{B}} \max_{S \subseteq \mathcal{I}} v_b(S)$, $\mathcal{A}$ takes as input an estimate of*

$$\max_{b \in \mathcal{B}} \max_{S \subseteq \mathcal{I}} (v_b(S) - \sum_{i \in S} \delta_i).$$

*Proof.* The algorithm $\mathcal{A}$ just uses $\mathcal{A}'$ to generate the price functions $\pi_i'()$, and sets $\pi_i(x) = \pi_i'(x) + \delta_i \mathbf{1}_{x>0}$. To show the social welfare guarantee for $\mathcal{A}$, we consider any sequence of buyers $\sigma = b_1, b_2, \ldots, b_m$ for which the optimal welfare is $W(\text{opt}(\sigma, \{c_i\}))$.

Below, we show how to construct another sequence of fake buyers $\sigma' = b_1', b_2', \ldots, b_m'$, and prove that

$$W(\text{opt}(\sigma, \{c_i\})) = W(\text{opt}(\sigma', \{c_i'\})) \tag{19}$$
$$W(\mathcal{A}(\sigma, \{c_i\})) = W(\mathcal{A}'(\sigma', \{c_i'\})) \tag{20}$$

Now the algorithm $\mathcal{A}'$ gives the guarantee that for all $\sigma'$,

$$W(\mathcal{A}'(\sigma', \{c_i'\})) \geq \frac{1}{\alpha}(W(\text{opt}(\sigma', \{c_i'\})) - \beta)$$

we would get the same guarantee for $\mathcal{A}$, and hence get the proof. The definition of the fake buyers $b_i'$ is natural: their valuation function is $v_i'(S) = v_i(S) - \sum_{i \in S} \delta_i$—note that fake buyers may have non-monotone valuation functions, and they may also have negative values for some sets, but this is not a concern. Now to prove (19) and (20).

**Claim D.2.** *For any $j \in [m]$, buyer $b_j \in \sigma$ buys the same set from $\mathcal{A}$ as $b_i' \in \sigma'$ buys from $\mathcal{A}'$.*

*Proof.* We prove this by induction. The base case is $j = 1$. Utility function $u_1(\cdot)$ for buyer $b_1$ is $\forall S \subseteq \mathcal{I}, u_1(S) = v_1(S) - \sum_{i \in S} \pi_i(1)$ while the utility function $u_1'(\cdot)$ for buyer $b_1'$ is $\forall S \subseteq \mathcal{I}, u_1'(S) = v_1'(S) - \sum_{i=1}^{S} \pi_i'(1)$. Using definition of $v_1'$ and $\pi_i$ we get

$$\forall S \subseteq \mathcal{I}, u_1'(S) = v_1'(S) - \sum_{i \in S} \pi_i'(1) = v_1(S) - \sum_{i \in S} c_i(1) - \sum_{i \in S} \pi_i'(1) = v_1(S) - \sum_{i \in S} \pi_i(1) = u_1(S).$$

Hence, $b_1$ and $b_1'$ have the same utility maximizing set and therefore they buy the same set of items.

Assume the induction hypothesis is true for $j < k$. We prove that buyer $b_k$ and $b'_k$ buy the same set of items. Since for all $j < k$, buyers $b_j$ and $b'_j$ bought the same set of items, therefore, the number of copies $x_i$ and $x'_i$ of item $i$ sold by $\mathcal{A}$ and $\mathcal{A}'$ when $b_k$ and $b'_k$ arrive are equal. Therefore,

$$\forall S \subseteq \mathcal{I}, \, u'_k(S) = v'_k(S) - \sum_{i \in S} \pi'_i(x'_i+1) = v_k(S) - \sum_{i \in S} c_i(1) - \sum_{i \in S} \pi'_i(x_i+1) = v_i(S) - \sum_{i \in S} \pi_i(x_i+1) = u_i(S).$$

Hence, $b_k$ and $b'_k$ buy the same set of items. This completes the step of induction. Hence proved. $\square$

Define an allocation vector $X_{ij} \in \{0,1\}^{n \times m}$ so that $X_{ij} = 1 \iff$ buyer $b_j$ is assigned a copy of item $i$.

**Claim D.3.** *Any allocation vector $X$ achieves equal social welfare on buyer sequence $\sigma$ with cost functions $\{c_i\}$, and on buyer sequence $\sigma'$ with cost functions $\{c'_i\}$.*

*Proof.* Denote by $y_i$ the number of copies of item $i$ allocated under the scheme $X_{ij}$; hence $y_i = \sum_j X_{ij}$. Also, let $S_j \subseteq U$ denote the set of items allocated to $j^{th}$ buyer.

Note that $W(X(\sigma, \{c_i\})) = \sum_j v_j(S_j) - \sum_{i \in \mathcal{I}} \sum_{k=1}^{y_i} c_i(k)$ and $W(X(\sigma', \{c'_i\})) = \sum_j v'_j(S_j) - \sum_{i \in \mathcal{I}} \sum_{k=1}^{y_i} c'_i(k)$. Using definition of $v'_i()$ and $\pi_i()$ we get

$$W(X(\sigma, \{c_i\})) = \sum_j v'_j(S_j) - \sum_{i \in \mathcal{I}} \sum_{k=1}^{y_i} c'_i(k)$$

$$= \sum_j \left(v_j(S_j) - \sum_{i \in S_j} c_i(1)\right) - \sum_{i \in \mathcal{I}} \sum_{k=1}^{y_i} c'_i(k)$$

$$= \sum_j v_j(S_j) - \sum_j \sum_{i \in \mathcal{I}} X_{ij} \cdot c_i(1) - \sum_{i \in \mathcal{I}} \sum_{k=1}^{y_i} c'_i(k)$$

$$= \sum_j v_j(S_j) - \sum_{i \in \mathcal{I}} y_i \cdot c_i(1) - \sum_{i \in \mathcal{I}} \sum_{k=1}^{y_i} c'_i(k)$$

$$= \sum_j v_j(S_j) - \sum_{i \in \mathcal{I}} \sum_{k=1}^{y_i} (c_i(1) + c'_i(k))$$

$$= \sum_j v_j(S_j) - \sum_{i \in \mathcal{I}} \sum_{k=1}^{y_i} c_i(k) = W(X(\sigma', \{c'_i\}))$$

which proves the claim. $\square$

Claim D.3 says having the same allocations in the two settings achieves the same social welfare; this proves (19). Moreover, by Claim D.2, the allocation made by $\mathcal{A}$ to buyer sequence $\sigma$ is the same as that made by $\mathcal{A}'$ to buyers $\sigma'$; this proves (20).

Finally note that in case $\mathcal{A}'$ needs an estimate of $\max_{b' \in \mathcal{B}} \max_{S \subseteq \mathcal{I}} v'_b(S)$ for its guarantee to hold, then $\mathcal{A}$ passes the estimate of $\max_{b \in \mathcal{B}} \max_{S \subseteq \mathcal{I}} (v_b(S) - \sum_{i \in S} \delta_i$ since by definition of $v'_b$ both quantities are equal. $\square$